\documentclass[pre,superscriptaddress]{revtex4}
\usepackage{amsmath,amssymb}
\usepackage{dcolumn}
\usepackage[dvipdfmx]{graphicx,color,hyperref}
\usepackage{wrapfig}
\usepackage{bm}
\usepackage{enumerate}
\usepackage{amssymb, amsthm, amsmath}
\usepackage{url}
\usepackage{ulem}
\newcommand{\NumR}{N_\mathrm{r}}

\newcommand{\NumF}{N_\mathrm{f}}
\newcommand{\NumS}{N_\mathrm{s}}

\newcommand{\Reads}{\mathbf{R}}

\newcommand{\BSeq}{\mathbf{B}}
\newcommand{\Sample}{\mathbb{S}}
\newcommand{\Types}{\mathbb{T}}
\newcommand{\Chr}{\mathbb{X}}

\newcommand{\freeSet}{\mathbf{C}}

\newcommand{\List}{\mathbb{L}}
\newcommand{\Memory}{\mathbb{M}}

\newcommand{\Cluster}{\mathbb{C}}

\newcommand{\chrX}{\mathrm{X}}
\newcommand{\chrY}{\mathrm{Y}}

\newcommand{\ra}{\mathrm{R}}
\newcommand{\hx}{h_{\gamma,x}}

\def\DNA{\mathop{\mathbb{D}}\nolimits}
\def\MQ{\mathop{\mathrm{MQ}}\nolimits}
\def\Bcell{\mathop{\mathrm{B }}\nolimits}
\def\CDfour{\mathop{\mathrm{CD4}^+ \mathrm{T}}\nolimits}
\def\CDeight{\mathop{\mathrm{CD8}^+\mathrm{T}}\nolimits}
\def\CLP{\mathop{\mathrm{CLP}}\nolimits}
\def\CMP{\mathop{\mathrm{CMP}}\nolimits}
\def\Ery{\mathop{\mathrm{Ery}}\nolimits}
\def\GMP{\mathop{\mathrm{GMP}}\nolimits}
\def\HSC{\mathop{\mathrm{HSC}}\nolimits}
\def\LMPP{\mathop{\mathrm{LMPP}}\nolimits}
\def\MEP{\mathop{\mathrm{MEP}}\nolimits}
\def\Mono{\mathop{\mathrm{Mono}}\nolimits}
\def\MPP{\mathop{\mathrm{MPP}}\nolimits}
\def\NK{\mathop{\mathrm{NK}}\nolimits}

\newcommand{\estim}[1]{\widehat{#1}}

\newcommand{\Mc}{M_{\rm cut}}
\newcommand{\re}{}
\newcommand{\rd}{}
\newcommand{\bl}{}
\newcommand{\gr}{}
\newcommand{\rdyi}{}
\newcommand{\ak}{}
\newcommand{\blue}{}
\newcommand{\rr}{}
\newcommand{\rrr}{}


\begin{document}
\title{Systematic clustering algorithm for chromatin accessibility data \\and its application to hematopoietic cells}


\author{Azusa Tanaka}
\affiliation{Department of Human Genetics, Graduate School of Medicine, The University of Tokyo}
\affiliation{Laboratory of Virus Control, Institute for Frontier Life and Medical Sciences, Kyoto University}

\author{Yasuhiro Ishitsuka}
\affiliation{Center for Science Adventure and Collaborative Research Advancement, Graduate School of Science, Kyoto University}
\affiliation{Department of Mathematics, Graduate School of Science, Kyoto University}

\author{Hiroki Ohta}
\affiliation{Center for Science Adventure and Collaborative Research Advancement, Graduate School of Science, Kyoto University}
\affiliation{Department of Physics, Graduate School of Science, Kyoto University}

\author{\\Akihiro Fujimoto}
\affiliation{Department of Human Genetics, Graduate School of Medicine, The University of Tokyo}
\affiliation{Laboratory of Virus Control, Institute for Frontier Life and Medical Sciences, Kyoto University}

\author{Jun-ichirou Yasunaga}
\affiliation{Laboratory of Virus Control, Institute for Frontier Life and Medical Sciences, Kyoto University}
\affiliation{Department of Hematology, Rheumatology and Infectious Disease, Faculty of Life Sciences, Kumamoto University}

\author{Masao Matsuoka}
\affiliation{Laboratory of Virus Control, Institute for Frontier Life and Medical Sciences, Kyoto University}
\affiliation{Department of Hematology, Rheumatology and Infectious Disease, Faculty of Life Sciences, Kumamoto University}


\date{\today}

\begin{abstract}
  The huge amount of data acquired by high-throughput sequencing requires data reduction for effective analysis.
  Here we give a clustering algorithm for genome-wide open chromatin
data using a new data reduction method.
This method regards the genome as a string of $1$s and $0$s based on
a set of peaks and calculates the Hamming distances between the strings.  
This algorithm with the systematically optimized set of peaks enables us to quantitatively evaluate differences between samples of hematopoietic cells and classify cell types, potentially leading to a better understanding of leukemia pathogenesis.
\end{abstract}

\maketitle


\section{Introduction}
Cellular phenotypes are governed by epigenetic mechanisms.
For example, \ak{information about how human DNA
  is packed and chemically modified in the nucleus plays an important role
  in understanding the differentiation and regulation of cells}
\cite{epigenome1,chromatin1,chromatin2,chromatin3}.
Methods such as chromatin immunoprecipitation sequencing (ChIP-seq) and assay for transposase accessible chromatin using sequencing (ATAC-seq) have proven useful for understanding the modification and detection of open chromatin on a genome-wide scale \cite{ATAC1, ATAC2, AML, CTCL, CLL}.
Those epigenetic data analysis methods usually start with data enrichment along the whole genome,
also known as ``peak calling'' \rr{\cite{QBreview,GBreview}}. 

Compared to RNA-seq data analysis, whose target regions are mainly in certain loci or genes across samples,
the target regions on epigenetic sequencing data are undetermined. To determine the target regions,
peak calling with an appropriate tool is often performed for the entire genome of every sample,
and the target regions are defined as merged peaks among all samples. Then the total number of reads or fragments present in each region
is counted for each sample, leading to a matrix, $X = (x_{i,j})$, where $x_{i,j}$ represents the number of reads/fragments from sample $i$ in region $j$.
The matrix elements are normalized by quantile normalization to reduce the biases arising from variations in the data size over samples,
followed by downstream processing \cite{AML,CTCL,CLL}.
 
However, this process raises two concerns. First, we do not fully understand the effect of merging all the peaks from different samples.
For example, if two peaks from different samples slightly overlap, those two peaks are considered as one peak after the peak merging step.
Therefore, the difference of the two peak positions, which may reflect cell identity, may be unintentionally ignored.
The second concern is that we have no justification for applying quantile normalization
over samples that are phenotypically different \cite{quantile,quantile2}.

Thus, the aim of the present study is to avoid these concerns by constructing an algorithm
that systematically classifies epigenetic data obtained from high-throughput sequencing.
In this analysis, toward cell type classification,
\rr{we provide a systematic algorithm to select a set of peaks used
for the downstream analysis}, where the difference between samples are quantified
by using the Hamming distance from information theory \cite{Hamming}.
\ak{This algorithm has less computational cost while still producing reasonable classification
  \rrr{compared to a previous method \cite{AML} }.}

\ak{As an application of the developed algorithm,
  we use it to obtain new insights
on samples of leukemia cells from chronic lymphocytic leukemia (CLL), acute myeloid leukemia (AML),
and adult T-cell leukemia (ATL) at the chromatin level.
In particular, using this algorithm, we infer the phenotype of a given leukemia sample
as output by using only ATAC-seq data of that sample as input.}

\section{Results}
\subsection{ATAC-seq samples} \label{Algo}
In this paper, we \ak{mainly} focused on $77$ ATAC-seq datasets from $13$ human primary blood cell types \cite{AML} as test data.
The $13$ cell types are comprised of hematopoietic stem cells (HSC), multipotent progenitor cells (MPP),
lymphoid-primed multipotent progenitor cells (LMPP), common myeloid progenitor cells (CMP), megakaryocyte-erythroid progenitor cells (MEP), granulocyte-macrophage progenitor cells (GMP), common lymphoid progenitor cells (CLP), natural killer cells (NK),
B cells, CD4${^+}$T cells (CD4${^+}$T), CD8${^+}$T cells (CD8${^+}$T), monocytes (Mono) and erythroids (Ery).
These cell types are experimentally categorized by immunophenotypes
described by the combination of cell surface markers shown in Table \ref{Table1}.
\begin{table}[htb]
\centering
  \begin{tabular}{|c|c|l|} \hline
    Cell type ($\nu$)&Number of replicates  & Immunophenotypes \\ \hline
    HSC      &7&  Lin-, CD34+, CD38-, CD10-, CD90+ \\ 
    MPP      &6&  Lin-, CD34+, CD38-, CD10-, CD90-\\
    LMPP     &3& Lin-, CD34+, CD38-, CD10-, CD45RA+\\
    CMP      &8&  Lin-, CD34+, CD38+, CD10-, CD45RA-, CD123+\\
    MEP      &7&  Lin-, CD34+, CD38+, CD10-, CD45RA-, CD123-\\ 
    GMP      &7&  Lin-, CD34+, CD38+, CD10-, CD45RA+, CD123+\\
    CLP      &5&  Lin-, CD34+, CD38+, CD10+, CD45RA+\\
    NK      &6&  CD56+\\
    B      &4&  CD19+, CD20+\\
    CD4$^+$T    &5&  CD3+, CD4+\\
    CD8$^+$T    &5&  CD3+, CD8+\\
    Mono    &6&  CD14+\\
    Ery     &8&  CD71+, GPA+, CD45-low\\
    \hline
  \end{tabular}
  \caption{
{\bf Immunophenotypes of samples.} 
    \rd{Types of hematopoietic cells and their corresponding cell surface markers in \cite{AML}.}
    For example, CD34+ and CD38- for cell type $\nu$ means that a cell of type $\nu$ expresses CD34
    \rd{but not CD38 at its surface.}
  }
  \label{Table1}
\end{table}

\re{For convenience, $\Types$ denotes a set of the thirteen cell types;}
\[
\Types = \{ \Bcell, \CDfour, \CDeight, \CLP, \CMP, \Ery, \GMP, 
	\HSC, \LMPP, \MEP, \Mono, \MPP, \NK\}.
\]

For all $77$ samples, we assigned ATAC-seq reads to reference genome hg19 (\url{http://hgdownload.cse.ucsc.edu/goldenPath/hg19/database/}),
and among them only those which had high mapping quality values (MQ $\geq$ 30)
were used for the peak calling by MACS2 \re{(see Appendix for details of the preprocessing)} \cite{MACS2}.
The peak calling results consisted of the location with a peak width and the associated $p$-value.
\re{Concretely, the location of the k-th peak is expressed by $g_k = (\gamma_k, \alpha_k, \beta_k)$, where
$\gamma_k$ is the chromosome number, $\alpha_k$ is the start position, and $\beta_k$ is the end position.}
Note that we used MACS2 to call all ATAC-seq peaks
with the following parameters ({-}{-}nomodel {-}{-}nolambda {-}{-}keep-dup all -p $p_G$),
where the number of peaks is affected by the peak calling parameter ``-p $p_G$''.
The parameter $p_G$ is larger than any $p$-values of the peak calling results.
\re{(See Materials and methods for details of the peak-calling.)}

\begin{figure}[htbp]
\includegraphics[width=13cm,clip]{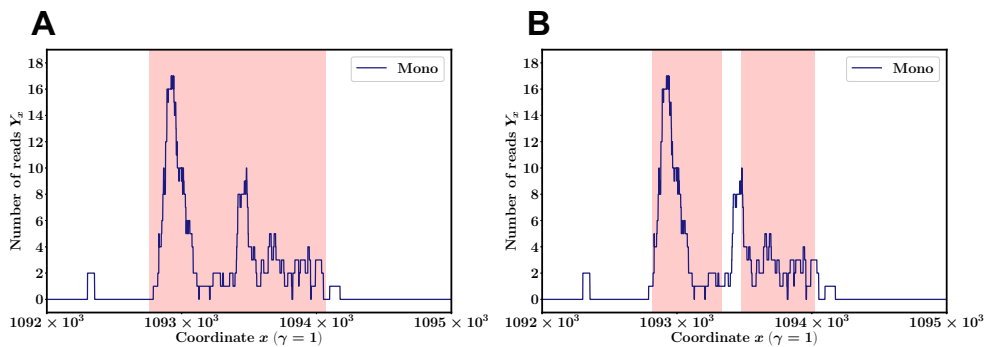}
\caption{
    {\bf The number of reads vs genomic positions.}
  The plots show representative data of Mono obtained
  from SRA with accession number SRR2920475.
  (A) The number
  of reads $Y_x$ at each position $x$ along chr $1$ ($\gamma = 1$) and the peak region $(\alpha_k,\beta_k)$
  as determined by the MACS2 algorithm with peak calling parameter $p_G=10^{-2}$ (pink shaded regions) is shown.  
  The peak region and its associated $p$-value $((\alpha_k,\beta_k),p_k)$ are $(1092756,1094068,10^{-20.36428})$. 
  (B) The obtained peak regions are $((1092817,1093330),10^{-20.36428} )$
  and $((1093480,1094025),10^{-8.19447})$ for $p_G=10^{-4}$.}
\label{Figure1}
\end{figure}

\ak{Note that the peak position depends on parameter $p_G$ of the MACS2 algorithm
as shown in Fig \ref{Figure1}. For example, the start and end positions of a peak could change
and one peak could split into two peaks depending on $p_G$.
Thus, we need to} \blue{take into account the dependence of a set of peaks on different values of $p_G$ for careful analysis.}

\subsection{Parameterized binarization} 
\begin{figure}[htbp]
\includegraphics[width=13cm,clip]{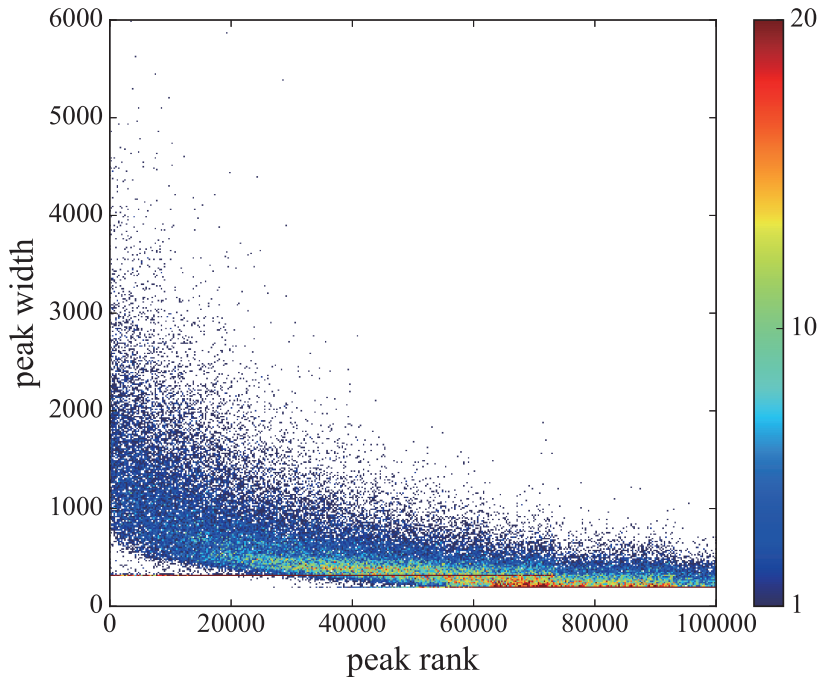}
 \caption{
    {\bf The statistics of peak width.}
    Distribution of peak width 
  $(\beta_k-\alpha_k)$  and its corresponding ranking $k$ 
  obtained from the peak calling result of CD4$^+$T cells with peak calling parameter $p_G=10^{-2}$.
  The bin size is $400\times 400$. \ak{The color code indicates the number of data in each bin.}}
\label{Figure2}
\end{figure}

First we ranked the peak results in the order of ascending $p$-values and then investigated the relationship
between the peak width and the corresponding ranking. 
We found that as the $p$-value increased, the width of the ATAC-seq peaks became shorter statistically,
which suggested the feasibility of \rr{robust} data reduction \rr{against small noise in the data}
by selecting peaks with smaller $p$-values (Fig \ref{Figure2}).

\begin{figure}[htbp]
\includegraphics[width=14cm,clip]{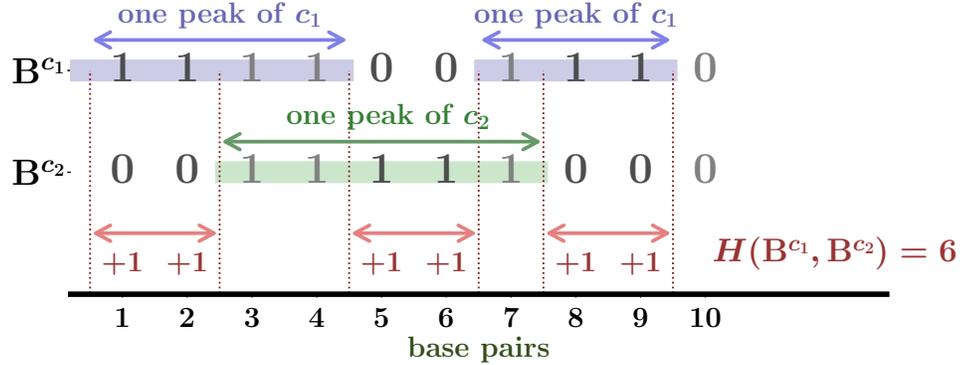}
\caption{
{\bf How to calculate Hamming distance.}
    Schema of the Hamming distance calculation from the peak locations
    \rr{with two samples $c_1,c_2\in\mathbb{S}$}.
    Each locus is converted to 1 or 0 based on the peak overlapping status.}
\label{Figure3}
\end{figure}

\begin{figure}[htbp]
\includegraphics[width=13cm,clip]{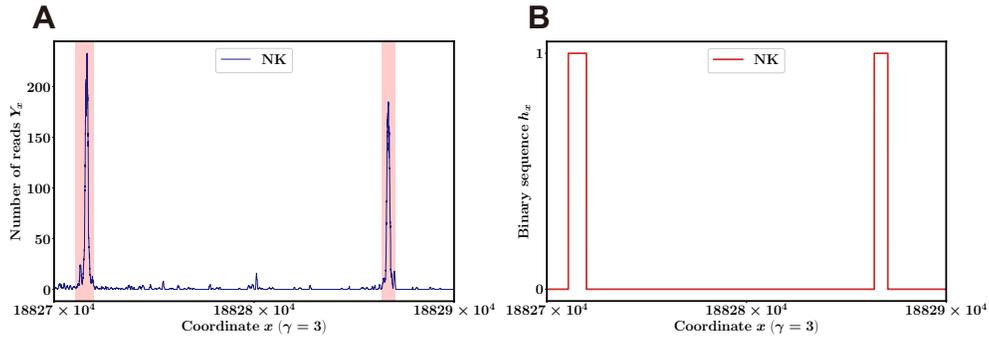}
 \caption{
{\bf Binarizing the number of reads.}
    (A) The number
  of reads $Y_x$ at each position $x$ along chr $3$ ($\gamma = 3$) and the peak region $(\alpha_k,\beta_k)$
  as determined by the MACS2 algorithm with peak calling parameter $p_G=10^{-2}$ (pink shaded regions).
  This figure shows representative data of NK cells obtained from SRA with accession number SRR2920495. 
  The peak regions and the associated $p$-values $((\alpha_k,\beta_k),p_k)$ in the \rd{left and right peaks 
  are $((188271079,188271985),10^{-422.5872})$ and $((188286401,188287077),10^{-329.52139})$, respectively.} 
  Thus, the width of the peaks $(\beta_k-\alpha_k)$ in the \rd{left- and right-hand sides} are $906$ and $676$, \rd{respectively}.
  (B) Binary sequence $(h_x)$ as determined by the peak regions seen in (A)
  when we chose $\Mc$ satisfying $p_{\Mc} \ge 10^{-329.52139}$.
}
\label{Figure4}
\end{figure}

Thus, we define $\Mc$ as the threshold such that only peaks with rankings not greater than
$\Mc$ are used for the analysis hereafter.
\rd{Then, for a given set of $(\Mc,p_G)$, we introduce $\mathbf{B}=\{h_{\gamma,x}\}$,
  where $h_{\gamma,x}=1$ when position $x$ in chromosome $\gamma$ 
  is inside a peak and $0$ otherwise (Fig \ref{Figure3}).} 
 The process to obtain the binary sequence from the reads data is illustrated in Fig \ref{Figure4}.
 Note that we do not perform any coarse-grained description for the genome position $x$
 but keep 1bp resolution.
 \re{(See Materials and methods for details of the binarization.)}

\subsection{Quantifying differences between two binary sequences by Hamming distance}\label{Distance}
Let us move onto the situation when one considers a set of samples to evaluate the difference
between two binary sequences $\mathbf{B}$. 
Here our strategy is to find the proper distance
that can be measured from the normalized ATAC-seq data of two samples.
Using that distance, we try to obtain hierarchical clustering of a set of hematopoietic cell samples
to quantitatively characterize the relationship among those samples.

Let $\NumS$ \rd{be the number of samples.
  \bl{We then write} the set of samples as
  \[
  \Sample:=\{1, 2, \dots, \NumS\},
\]
}
where $\NumS = 77$ in this study.
For sample $c \in \Sample$, we add index $c$ to related objects as a superscript.
For example, we write a binary sequence $\mathbf{B}$ associated to sample $c$ as $\mathbf{B}^c:=\{\hx^c\}$.

\begin{figure}[htbp]
\includegraphics[width=10cm]{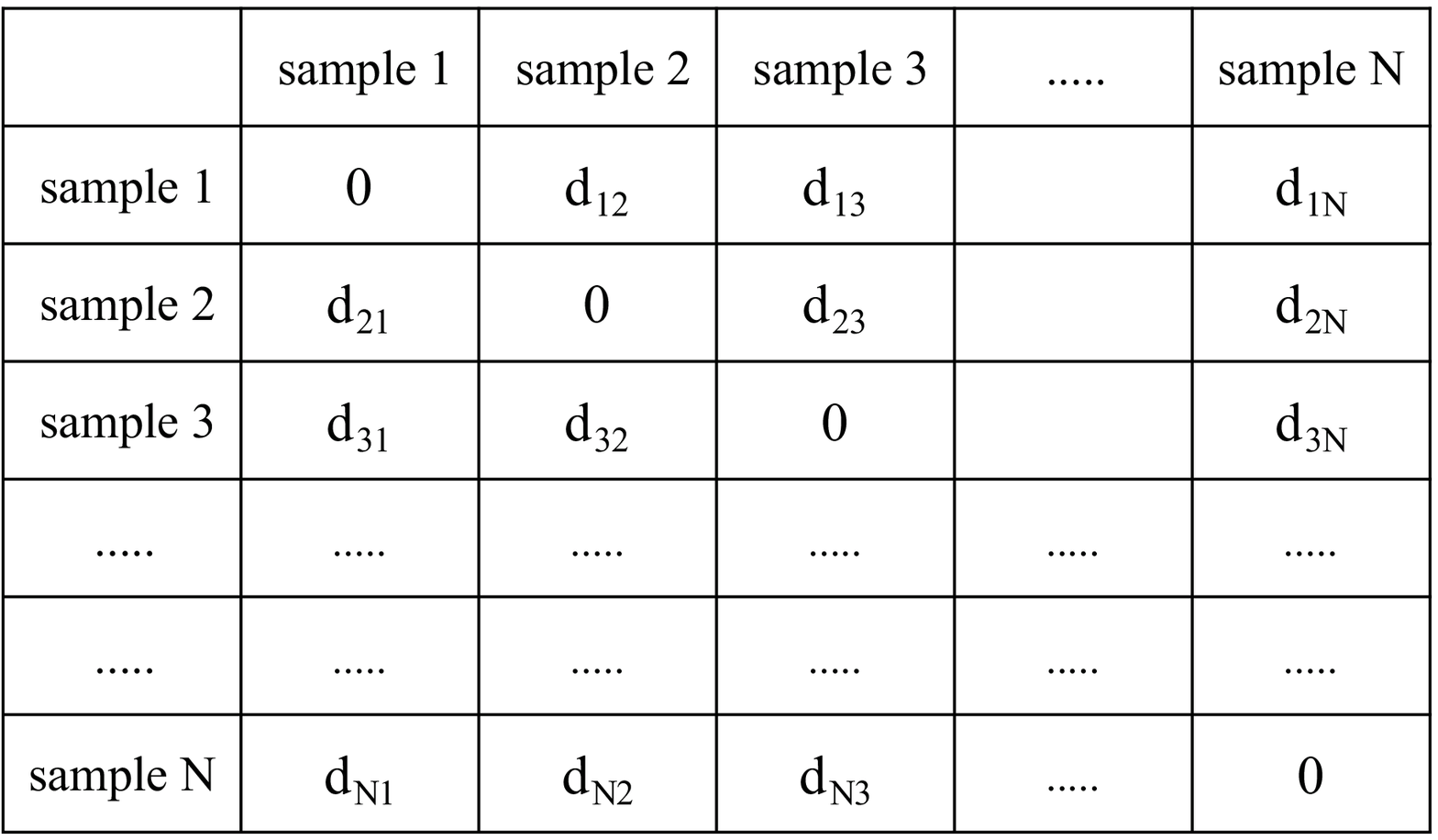}
\caption{
{\bf Matrix of Hamming distances.}
    Matrix of Hamming distances $d_{ij}$ between samples $i$ and $j$.
  This matrix is used for the downstream analysis.}
\label{Figure5}
\end{figure}

\begin{figure}[htbp]
\includegraphics[width=16cm,clip]{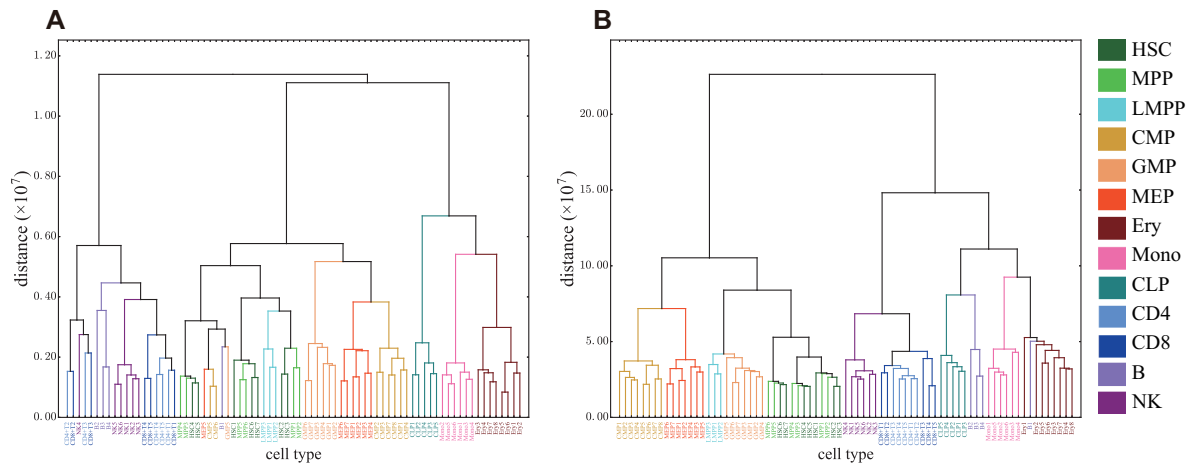}
  \caption{
{\bf Examples of clustering dendrograms. }
    Hierarchical clustering obtained by Ward's method
    with parameters  $(\Mc,p_G)=(2000,10^{-2})$
    (A) \gr{and} $(80000,10^{-2})$ (B).
}
\label{Figure6}
\end{figure}

There are many methods to evaluate the difference between
a binary sequence $\mathbf{B}^c$ from sample $c\in\mathbb{S}$ and
$\mathbf{B}^{c'}$ from sample $c'\in\mathbb{S}$.
In this paper, we evaluated the difference between two samples $(c,c')$ by using the Hamming distance $H(\mathbf{B}^{c}, \mathbf{B}^{c'})$
between two binary sequences, $\mathbf{B}^{c}$ and $\mathbf{B}^{c'}$. 
\re{$H(\mathbf{B}^{c}, \mathbf{B}^{c'})$ is calculated as the sum of the number of pairs with different values at every position $x$
between $\mathbf{B}^c$ and $\mathbf{B}^{c'}$ (Fig \ref{Figure5}).}
We used the distance as an initial condition for the hierarchical clustering
and then used \ak{Ward's method} to complete the hierarchical clustering \cite{Hier}.
\gr{Examples of} hierarchical clustering with $(\Mc,p_G)=(2000,10^{-2})$ \gr{and} $(80000,10^{-2})$
are shown in Fig \ref{Figure6}.
\re{(See Materials and methods for details of the Hamming distance and hierarchical clustering.)}

\subsection{Optimization of hierarchical clustering \rr{toward} cell-type classification} \label{Opt}
By using the methods explained above, we can obtain a clustering dendrogram \gr{that depends on $(\Mc, p_G)$}.
We then need to systematically determine the best clustering
\ak{, which is the clustering closest to the ``perfectly classified dendrogram'' where}
\blue{each} set $\mathbb{S}_\nu$ of all samples with type $\nu\in\Types$ coincides with an offspring set.
This \rd{condition} can be restated as an optimization problem 
by introducing a cost function ``penalty'' for the performance of clustering \ak{as follows}.

\begin{figure}[htbp]
\includegraphics[width=12cm,clip]{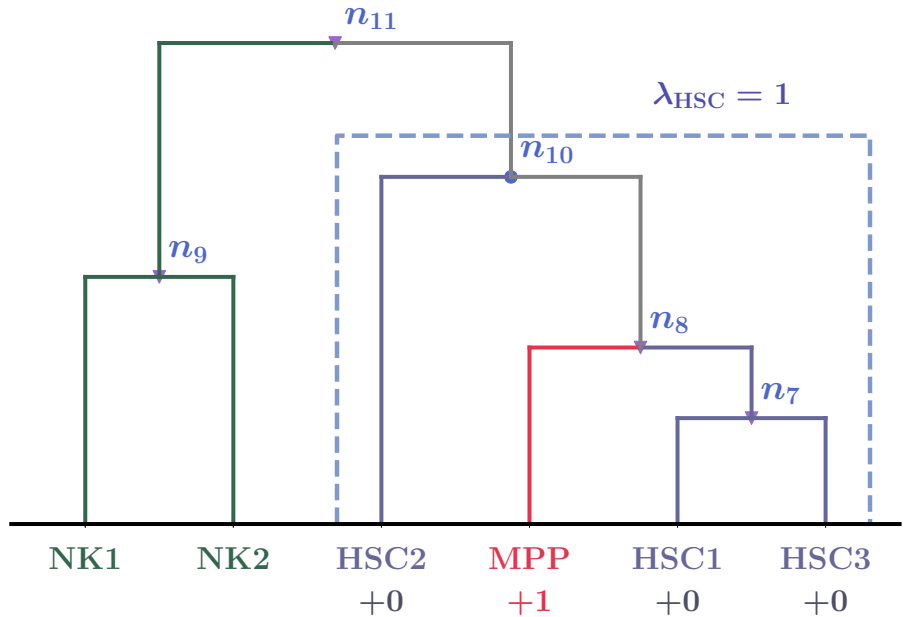}
\caption{
  {\bf Schema of penalty score calculation.}
   Note that this dendrogram is constructed by artificial data to explain how to calculate the penalty,
   though we use the same labels such as HSC1.
This dendrogram has six leaves, and three of them are classified to type $\HSC$.
To explain details of this dendrogram, we freely use the symbols and definitions
in Materials and methods in this caption.
We can see that $\tau(\HSC) = 10$. The corresponding node is $n_{10}$ (displayed by the blue dot),
and the corresponding cluster $\Cluster_{10}$ is the set $\{\HSC1, \HSC2, \HSC3, \MPP\}$
(surrounded by the blue \rr{dashed} line).  
Among \rd{the elements of  $\Cluster_{10}$, one leaf, $\MPP$, is not in type $\HSC$, but the three others are.} 
Hence, the type penalty of $\HSC$ in this figure is computed as
\(
	\lambda_{\HSC} = 4 - 3 = 1.
\)
}
\label{Figure7}
\end{figure}

Concretely, to quantitatively evaluate the obtained dendrogram for each combination of \re{$(\Mc, p_G)$},
we define type penalty $\lambda_\nu$ for a given cell type $\nu\in\Types$.
Type penalty $\lambda_\nu$ corresponds to the number of samples from different cell types in cluster
$\nu$ formed when all samples of cell type $\nu$ meet together \re{from the bottom of}
the dendrogram (Fig \ref{Figure7}).
Additionally, we define \rr{global} penalty \ak{$\lambda:=\sum\limits_{\nu\in \mathbb{T}} \lambda_\nu$} as
the ``cost function'' of the optimization.
\blue{Note that $\lambda\ge 0$, and a ``perfectly classified dendrogram'' gives $\lambda=0$.}
(See \re{Materials and methods} for details of the penalty.)

\subsection{Determination of the best parameters \rr{for the optimization}} \label{Bestpara}
\begin{figure}[htbp]
\includegraphics[width=13cm,clip]{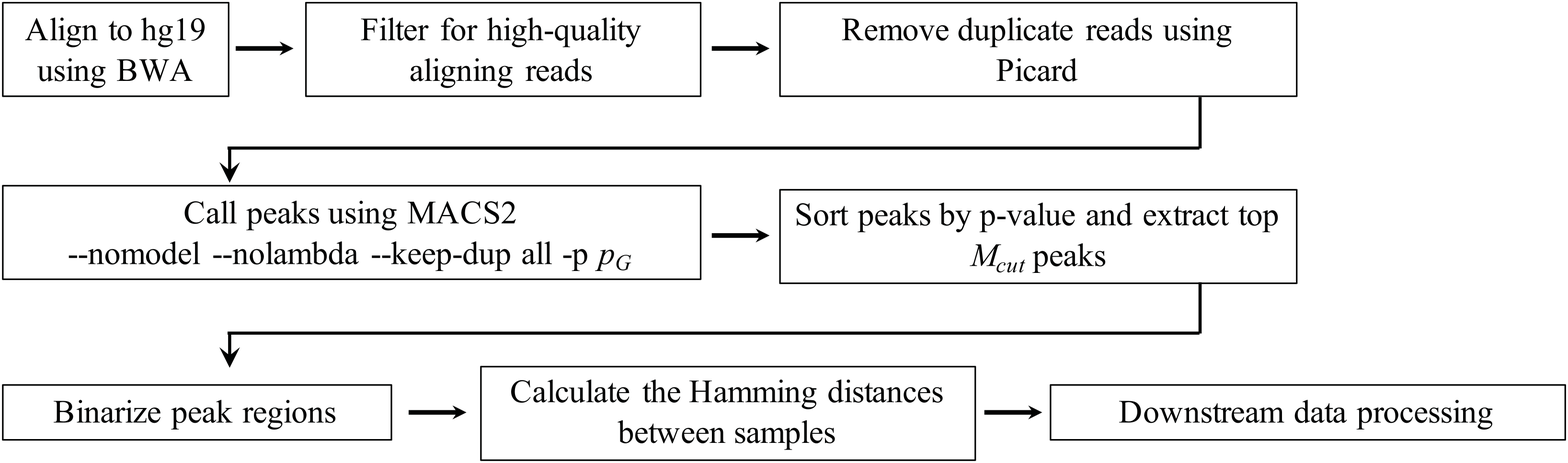}
\caption{
  {\bf Schematic workflow of our algorithm.} 
  See Materials and methods for details.}
\label{Figure8}
\end{figure}

\re{As mentioned above,}
the optimization problem we have to solve is \rd{to find $(\Mc^*,p_G^*)$
  that minimizes the cost function} $\lambda(\Mc,p_G)$.
\rr{The schematic workflow in our algorithm is shown in Fig \ref{Figure8}.}

\begin{figure}[htbp]
\includegraphics[width=10cm,clip]{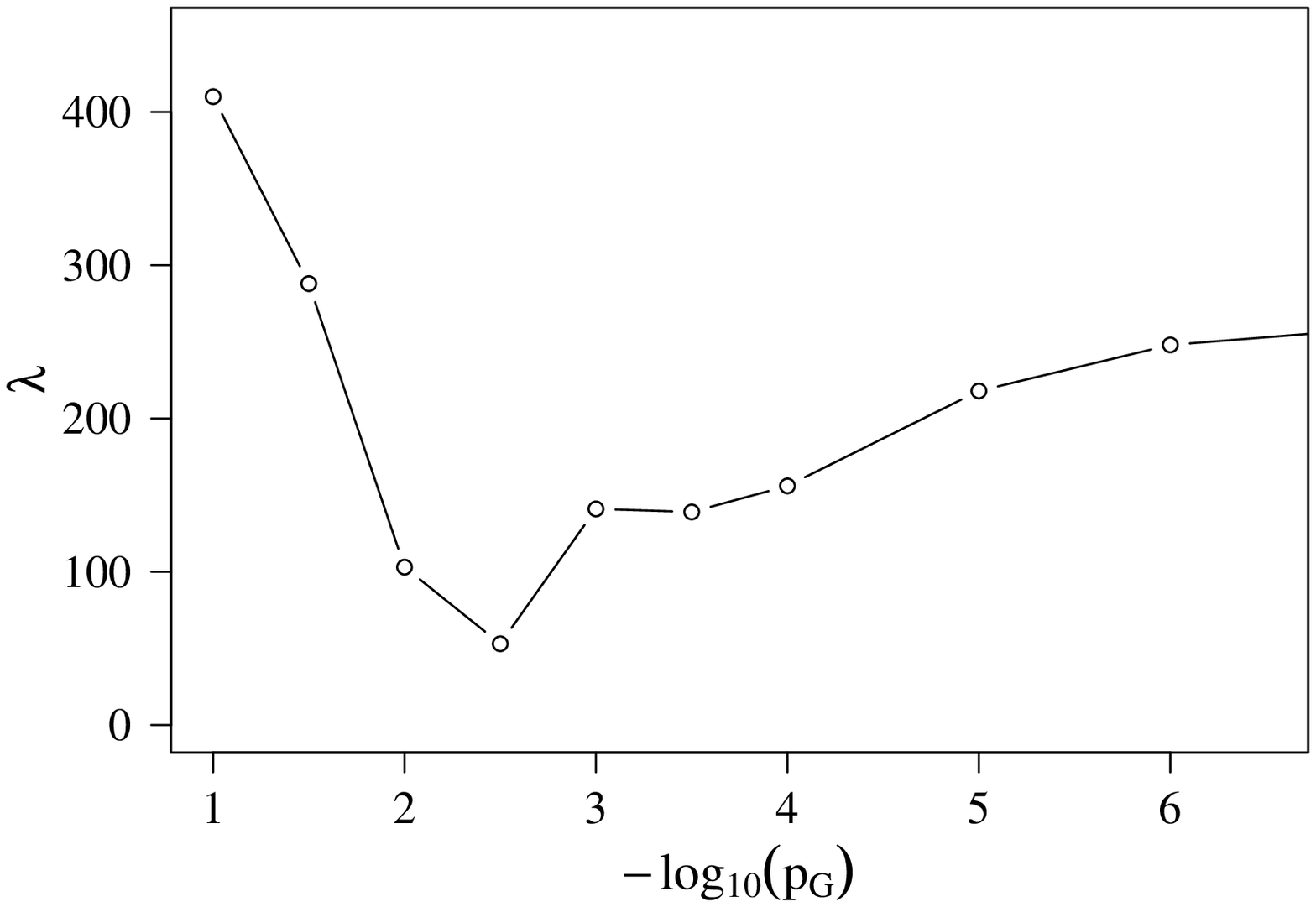}
\caption{
{\bf Global penalty without cutoff of reads.}
  Global penalty $\lambda (\Mc=\infty,p_G)$ obtained by Ward's method.}
\label{Figure9}
\end{figure}

\rd{First we took into account all the peaks by setting $\Mc =\infty$
  and checked how the dendrograms and $\lambda(\infty,p_G)$
  depended on $p_G$, as shown in Fig \ref{Figure9}. Considering the \ak{tendency} of the parameter searching,
  we concluded that $1.5\le -{\log_{10}p_G^*}\le 4$.}

\begin{figure}[htbp]
\includegraphics[width=16cm,clip]{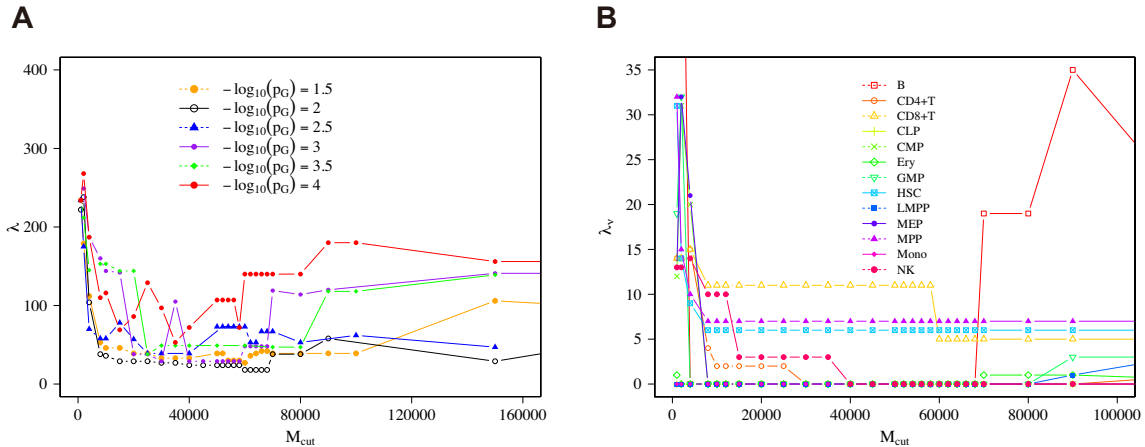}
  \caption{
{\bf Penalty with cutoff of reads. }
    The distribution of global penalty $\lambda$ (A) and type penalty $\lambda_\nu$
    for each cell type $\nu$ (B) along with $\Mc$ with parameter $p_G=10^{-2}$ by Ward's method. }
\label{Figure10}

\end{figure}

\begin{figure}[htbp]
\includegraphics[width=16cm,clip]{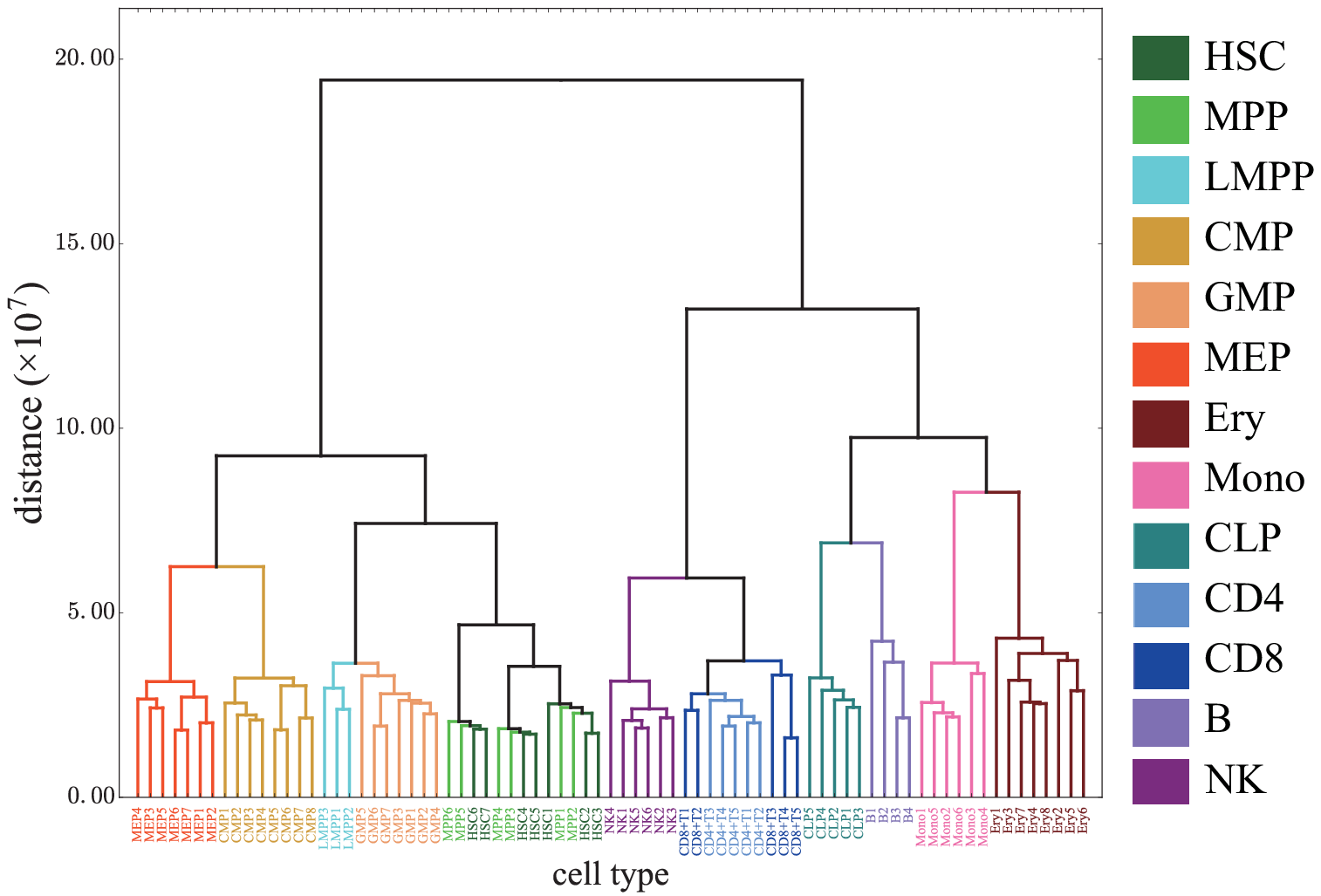}
  \caption{ {\bf Our best clustering dendrogram.}
    Hierarchical clustering obtained by Ward's method with $(\Mc,p_G)=(64000,10^{-2})$.}
\label{Figure11}
\end{figure}

We then sought the best parameters to optimize the dendrograms and found that $(\Mc^*,p_G^*)$
was close to $(64000,10^{-2})$, which gave the smallest penalty $\lambda$
in our searching resolution, as shown in Figs \ref{Figure10} and \ref{Figure11}.
\rr{Note that $64000$ is the midpoint of $(60000,62000,64000,66000,68000)$
  which give the same minimum penalty in our searching resolution.}
Hereafter, to investigate the property of the best clustering,
we \rd{set} $(\Mc^*,p_G^*)$ as $(64000,10^{-2})$.
In our searching resolution, \rr{the increment in terms of $\Mc$ was $2000$} near $\Mc=64000$.
\rr{Note that more-refined resolutions might give better estimates of the optimized value $(\Mc^*,p_G^*)$,
but naturally the computational costs get higher.
Even then, the following procedures are operationally unchanged.}

 \rr{The value of the minimum penalty achieved at $(\Mc^*,p_G^*)$ was $18$}.
  This minimum was smaller than the penalty value of $27$ for the clustering of the data
  from GSE74912\_ATACseq\_All\_Counts.txt in \cite{AML}.
  The procedure of the latter clustering was as follows.
  First we performed a quantile normalization of the reads count in the distal elements
  ($>$ 1000 bp away from a transcription start site (TSS)).
  Then we calculated the Pearson coefficients over all samples
  leading to a distance matrix where each entry is 1-(Pearson coefficient).
  By using Ward's method, we finally obtained the clustering dendrogram.
  \rr{Note that for this case, Ward's method gives penalty $\lambda=27$ and UPGMA gives $\lambda=29$.}

\subsection{Computational cost of the algorithm} \label{Compcost}
As explained above, after obtaining data of the reads positions,
we perform the MACS2 algorithm to get peak regions,
and then finally we produce a hierarchical clustering.
Here we consider the computational cost of our algorithm
after acquiring \rr{the data of the reads positions}
and until acquiring a distance matrix to produce the hierarchical clustering.
\rr{Note that the computational cost of the MACS2 algorithm is not more than $O(\NumS)$,
where $O()$ is the Landau notation and $\NumS$ is the total number of samples.}
We consider two situations. (i) One is the case \rr{where new samples to analyze are given.}
(ii) The other is the case where one new sample to analyze is added to the already analyzed samples,
for which peak regions and the distance matrix are already calculated.
\rr{For case (ii), we use the symbol $\NumS$ to write the total number
of already analyzed samples.
We claim that the computational cost of our algorithm is
significantly lower}
than that of \rrr{a previous method} using target regions merged over samples \cite{AML}
\rr{for large values of $\NumS$ for case (ii)
  and, in our case with $\NumS$=77,
  that the computational cost of our algorithm is practically lower for case (i)}.

Specifically, in case (i) for our algorithm, the corresponding computational cost
  is \rr{$K_1\Mc\NumS^2$}, which comes solely from the calculation of the Hamming distance.
  In case (ii), the corresponding computational cost is \rr{$K_2\Mc\NumS$},
  which also comes solely from the calculation of the Hamming distance.
  \rr{Note that $K_1$ and $K_2$ are constants that do not depend on $\Mc$ or $\NumS$.}

In the context of estimating the best optimization parameter $\Mc^*$,
by using $M_{\mathrm m}$ different values for $\Mc$,
the computational cost becomes \rr{ $K_1\Mc M_{\mathrm m}\NumS^2$} for case (i) and
\rr{$K_2\Mc M_{\mathrm m}\NumS$} for case (ii), where $M_{\mathrm m}$ does not depend on $\NumS$ or
genome size $L$ and can be adjusted according to the searching resolution of the optimization.
\rr{Note that $K_1$ and $K_2$ do not depend on $M_{\mathrm m}$.}
In addition, we optimize $p_G$ by $M_{\mathrm p}$ different values for $p_G$.
Since this optimization can be done for any algorithm,
we do not take into account this cost for the comparison of different algorithms.
Typically, we set $(M_{\mathrm m},M_{\mathrm p})\simeq(30,10)$ in our optimization corresponding to case (i).
  Note that in the section of ``Application to leukemic cells'' discussed later, corresponding to case (ii),
  we use the optimized parameters $(\Mc,p_G)=(\Mc^*,p_G^*)$, leading to $(M_{\mathrm m},M_{\mathrm p})=(1,1)$.

  The \rrr{previous method} using targeted regions merged over samples in \cite{AML} includes
  (a) \rr{the merging of reads before peak calling}
    and (b) calculating the distance matrix by the Pearson coefficients
    which automatically depend on $\NumS$. 
  Thus, \rr{for a given number $N_{\rm new}$ of unanalyzed samples,
  the computational cost corresponding to the process of (a) and (b)
  is  at least $K_rN_rN_{\rm new} + K_LL_1\NumS^2$},
  where $N_r$ is the minimum reads number over all samples,
  and  $L_1$ is the number of target regions merged over all samples.
  The first term comes from counting the reads
  and the second term comes from calculating the distance matrix.
  \rr{Note that $K_r$ is a constant that does not depend on $N_r$ or $N_{\rm new}$,
  and $K_L$ is a constant that does not depend on $L_1$ or $\NumS$.}
  This form of the computational cost \rr{$K_rN_rN_{\rm new} + K_LL_1\NumS^2$}
  is the same for case (i) with $N_{\rm new}=\NumS$ and case (ii) with $N_{\rm new}=1$,
  leading to the conclusion that the computational cost of our algorithm is
  significantly lower than the \rrr{previous method}, especially for case (ii)
  \rr{with sufficiently large $\NumS$}.
  \rr{We do not have the exact estimate of the coefficients $K_1,K_2,K_r,K_L$,
   but because}
 $N_r=3265006 \gg \Mc^*$ and $L_1=590650 \gg \Mc^*$ in our case,
  \rr{then $K_rN_rN_{\rm new} + K_LL_1\NumS^2$
    could be costly compared to $K_1\Mc\NumS^2$.}
  In practice, even in case (i) \rr{with $\NumS=77$,
    we numerically found that the computational cost of}
  our algorithm is \rr{lower due to our algorithm not using the process of
    merging reads unlike \cite{AML}.}

\subsection{How to relate the best parameters to genomic context} \label{parageno}
In order to understand why ATAC-seq data under the condition of
$(\Mc, p_G) = (64000, 10^{-2})$ was well classified,
we analyzed the properties of the peaks with higher rankings.

\gr{The result of the previous section suggested that} \rd{peaks of $\{g_k\}_{k=1}^{\Mc^*}$
  with $\Mc^*=64000$ included key regions for characterizing} cell types.
Therefore, we investigated \ak{which functional genomic regions such as promoters, enhancers, etc.
  are dominantly related to these top $64000$ peaks.}

\subsubsection*{Functional annotation of peaks depending on rank}\label{Renhancer}
In order to investigate functional annotations on the genome
 overlap with \rd{ATAC-seq peaks data},
 we \rd{applied} the top 80000 peaks in three cell types (HSC, B cells, and Mono)
 to the 15-state ChromHMMmodel data.
One can obtain \rd{data of the biological functions on the genome for HSC, B cells, and Mono
  from an integrative analysis of \bl{111 reference human epigenome datasets},} 
where we used the data of E032 for B cells, E035 for HSC, and E029 for Mono (https://egg2.wustl.edu/roadmap/data
/byFileType/chromhmmSegmentations/ChmmModels/coreMarks/jointModel/final/) \cite{chromhmm1}.

\begin{figure}[htbp]
 \includegraphics[width=18cm,clip]{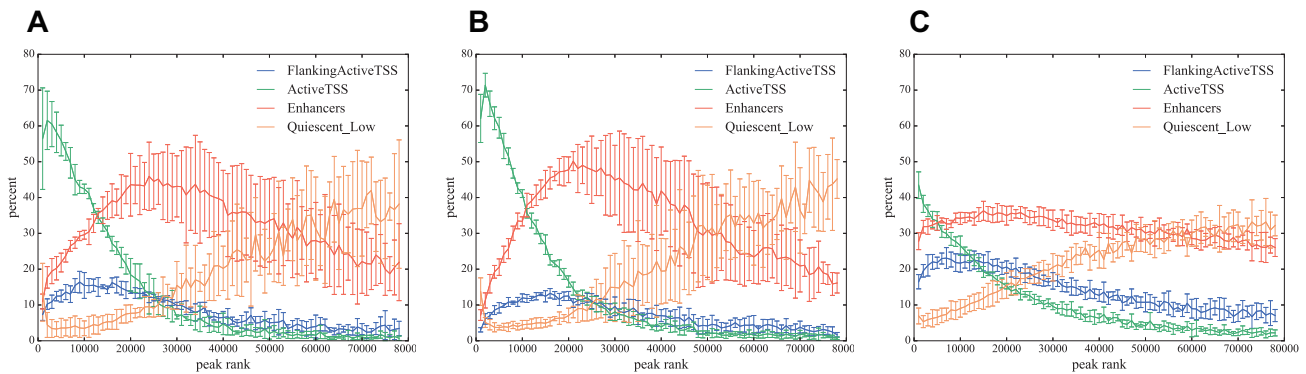}
  \caption{
    {\bf Functional annotations of peaks.}
    Percentage ($100\times |E_x^y|/1000$) of functional annotations 
    in every $1000$ peaks for B cells (A), Mono (B), and HSC (C).
    Only the functional annotations that have maximum percentages $\ge 12\%$,
    $y\in\{\rm FlankingActiveTSS, ActiveTSS, Enhancers, Quiescent\_Low\}$,
    are shown.}
\label{Figure12}
\end{figure}

ATAC-seq peaks were ranked according to $p$-values and divided into groups consisting of 1000 peaks.
Then we calculated the average ratio and the standard deviation for each of the 15 states over all samples in each cell type. For an explicit description,
let us introduce a set of functional annotations, $\mathbb{W}:=\{\mathbb{W}_y\}_{y=1}^{15}$,
where $\mathbb{W}_y$ is the set of \gr{regions} on the genome, each of which corresponds to functional annotation $y$.
\bl{We want to know how many peaks, $k$, of every 1000 peaks 
belong to each functional annotation $y$.}
For this purpose, we define
\begin{eqnarray*}
  E_x^y:=\left\{
  x\le k < x+1000 \ \middle|\mathrel{}
  \
  \exists( \gamma_k,[\sigma,\epsilon] ) \in \mathbb{W}_y \ such\ that\ \sigma\le (\alpha_k+\beta_k)/2\le \epsilon
\right\},
\end{eqnarray*} where $g_k=(\gamma_k,\alpha_k,\beta_k)$ is the peak position.
We computed $|E_x^y|/1000$ for $x\in\{1+(j-1)\times1000\}_{j=1}^{80}$, as shown in Fig \ref{Figure12}.
Note that we used the position of the peak center, ($\alpha_k+\beta_k)/2$, to annotate biological function.

As shown in \bl{Fig \ref{Figure12}}, most of the peaks with higher rankings \gr{belonged}
to \gr{``Active TSS'', which was related to the promoters of active genes},
but as the rank went down, the ratio of peaks from enhancer regions started to increase.
\ak{As the rank went down further, the ratio of peaks from ``quiescent-low'' regions started to increase.
The ratio of peaks from promoters and enhancers crossed at around peak rank 10000
and the ratio of peaks from enhancers and ``quiescent-low'' regions crossed at around peak rank 60000.}
\ak{Therefore, we concluded that the number around the $64000$th peak
is strongly related to the point that the contribution of ``quiescent-low'' regions to the Hamming distances
exceeds the contribution of enhancer regions to the Hamming distances.}

  Note that the \rr{type penalty} of HSC under the condition $(\Mc^*, p_G^*)$
was not as good as that of B cells or Mono,
and the functional annotation result of HSC did not show clear behaviors compared \rd{with} B cells
and Mono (Fig \ref{Figure12}C), which may partially explain
the worse \rr{type penalty} of HSC (Fig \ref{Figure10}B).

\subsection{Variations of hierarchical clustering methods} \label{Rob}
In general, \rd{when one performs data clustering,
  the effect \gr{of variations of} the clustering algorithms
  and the effect of loss of data on the clustering output should be considered.}

\begin{figure}[htbp]
 \includegraphics[width=16cm,clip]{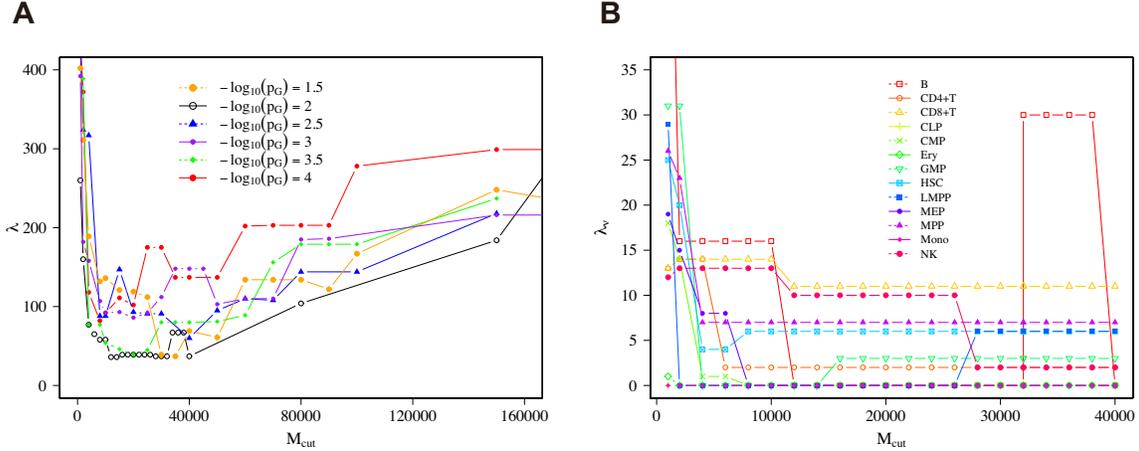}
\caption{
{\bf Penalty by UPGMA method.}
      The distribution of global penalty $\lambda$ (A) and type penalty $\lambda_\nu$
    for each cell type $\nu$ (B) along with $\Mc$ with parameter $p_G=10^{-2}$ by using UPGMA. }
\label{Figure13}
\end{figure}

\rd{First we considered \gr{the dependence of the clustering results
    on the variations of the clustering algorithms}.
  \gr{Besides Ward' method} which we used until here,
  there are several hierarchical clustering methods including UPGMA (Unweighted Pair Group Method with Arithmetic mean),
  WPGMA (Weighted Pair Group Method with Arithmetic Mean),
  UPGMC (Centroid Clustering or Unweighted Pair Group Method with Centroid Averaging),
  and WPGMC (Median Clustering or Weighted Pair Group Method with Centroid Averaging).
  We performed optimization also with UPGMA, as shown in Fig \ref{Figure13},
  and found that the minimum value of the penalty is $36$ with $\Mc=12000$.
  The other methods give worse results in general.
  Specifically, the minimum values of the penalty we found were
  $59$ for WPGMA with $\Mc=20000$,
  $127$ for UPGMC with $\Mc=30000$, and
  $149$ for WPGMC with $\Mc=35000$.
  These results suggested that Ward's method giving $18$ as the minimum value of the penalty
  was a better choice than that of the other methods for our purpose.}

\subsection{Robustness of our best clustering against the loss of data}
\begin{figure}
\includegraphics[width=16cm,clip]{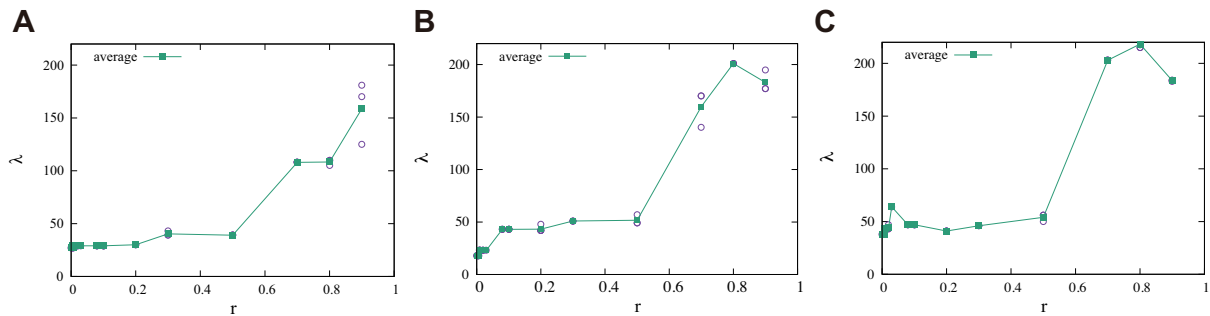}
 \caption{
{\bf Robustness of penalty against the loss of reads data.}
    The effect of the loss of reads on the global penalty $\lambda$.
    Reads were removed randomly from the uniform distribution with probability $r$.
  Then global penalty $\lambda$ was calculated with parameter
  $\Mc=30000$ (A), $\Mc=64000$ (B) or $\Mc=80000$ (C).
  Each circle indicates one sample and each square indicates the average over samples at the same $r$ value.}
\label{Figure14}
\end{figure}

\rd{Regarding the loss of data, let us consider making new reads data $\hat{\Reads}$ from original data $\Reads$.
  Specifically, we set $r$ with $0\le r\le 1$ as the probability of randomly removing  $\lceil r \NumR \rceil$
  reads from $\Reads$ with the uniform distribution, where $\lceil \chi \rceil$ means}
\gr{the minimum integer larger than or equal to $\chi$}.
Thus we can obtain $\hat{\Reads}=\{\Reads_i'\}_{i=1}^{\NumR-\lceil r \NumR \rceil}$,
where $\Reads_i'$ is one read in $\Reads$.
\re{Using this procedure,} we computed $\lambda$ for $(\Mc^*,p_G^*)=(64000,10^{-2})$.
As shown in Fig \ref{Figure14}B,
\ak{when ratio $r$ was increased, the value of $\lambda$ was constant until $r=0.007$
and gradually increased thereafter. In the region $r \geq 0.7$, $\lambda$ increased dramatically.
Note that $r=0$ gave $\lambda=18$ and the highest possible value of $\lambda$ for $77$ samples is $924$.
Thus, we concluded that for small $r$, the average penalty tended to be stably close to that of $r=0$.}

\ak{Further, we investigated $\lambda$ for different values of $\Mc$ than $64000$
to check the robustness of $\Mc^*$ against random selections.
Specifically, we investigated the behavior of $\lambda$ by varying $r$
  for $\Mc=30000$ and $80000$ with $p_G^*=10^{-2}$.
  The minimum value of $\lambda$ as a function of $r$ was $27$ for $\Mc=30000$ and located at $r=0$ (Fig \ref{Figure14}A)
  and was $38$ for $\Mc=80000$ and again located at $r=0$ (Fig \ref{Figure14}C).
  Note that in the region $r\ge 0.08$, $\lambda$ for $\Mc=30000$ was smaller than $\lambda$ for $\Mc=64000$,
  which suggested that $\Mc^*$ becomes less than $64000$ when the data size is decreased.}

\blue{Thus, for the present data size, we concluded that our algorithm was stable against small losses of the data 
  and it could also work well by adjusting $\Mc$ for losses of data up to $50$ percent.
 The obtained results imply that when the given data size is increased, our algorithm becomes more stable 
or potentially achieves better clustering with a smaller penalty than our current best clustering. }
  
\section{Application to leukemic cells} \label{App}
To evaluate the practicality of our algorithm \rr{with the optimized parameters $(\Mc^*,p_G^*)$} on cancer research,
we analyzed \blue{three types of leukemia:
CLL, AML, and ATL,}
by calculating \ak{Ward's distance function, $\mathcal{H}_{\mathrm{Ward}}(\zeta,\mathbb{S}_\nu)$,
between a given leukemia sample $\zeta$ and all samples $c\in\mathbb{S}_\nu$ of cell type $\nu$.
(See Materials and methods for details of $\mathcal{H}_{\mathrm{Ward}}$.)}

\blue{To separate normal and leukemic cells effectively, information about the cell surface markers was used. 
CLL is a disease that is characterized by the clonal proliferation of malignant B lymphocytes. Leukemic cells from CLL patients were purified by using the cell surface markers CD5 and CD19, which are commonly used as markers for CLL (Table \ref{Table2}) \cite{CLL2020}.
}

\begin{table}
\begin{center}
  \begin{tabular}{|l|l|}
    \hline
    Type of sample & \rd{Marker expression}\\ \hline
    CLL &  CD19+, CD5+\\
    \hline
    AML pHSC &  Lin-, CD34+, CD38-, TIM3-, CD99-\\
    AML LSC & Lin-, CD34+, CD38-, TIM3+, CD99+\\
    AML Blast &  Non-LSC; CD45-Intermediate, SSC-High\\
    \hline
      ATL &  CD4+, CADM1+ \\
    \hline
  \end{tabular}
  \end{center}
\caption{
{\bf Immunophenotypes of leukemic samples.}
\\\re{Immunophenotype of CLL} \cite{CLL}:
  Note that B cells are CD19+, as shown in Table \ref{Table1}.\\
\rr{\re{Immunophenotype of AML} \cite{AML}: SSC-high means that the intensity of side scatter in the flow cytometry is high.
Note that HSC, MPP, and LMPP are Lin-, CD34+, CD38- as shown in Table \ref{Table1}.\\
\re{Immunophenotype of ATL} \cite{cadm1,cadm1_2}:
  Note that CD4$^+$T cells are CD4+, as shown in Table \ref{Table1}}.}
  \label{Table2}
\end{table}

The AML samples analyzed in this study were divided into three stages, preleukemic HSC (pHSC), leukemia stem cells (LSC), and AML blasts by cell surface markers according to \cite{AML} (Table \ref{Table2}).
Briefly summarizing these three types, HSC that acquired founder mutations become pHSC, which expand to generate preleukemic clones. 
The subsequent acquisition of progressor mutations creates LSC, which can self-renew and produce AML blasts\cite{AMLreview}.
\blue{It has been reported that mature LSC populations more closely resemble normal GMP, and immature LSC populations are functionally similar to LMPP \cite{LSCreview}.
  A recent study has revealed that CD99-positive cells are almost entirely
  composed of LMPP-like cells in the sense of Ref. \cite{Chungeaaj2025}.
  Thus, the LSC used in our study, which are CD99-positive, can be presumed to be LMPP-like LSC.}

\blue{Human T-cell leukemia virus type 1 (HTLV-1) is a causative agent of ATL and HTLV-1-associated myelopathy/tropical spastic paraparesis (HAM/TSP)  \cite{ATL}. ATL has been subclassified into four clinical subtypes: acute, lymphoma, chronic, and smoldering. The chronic and smoldering subtypes are considered indolent, while patients with the acute or lymphoma subtype generally have a poor prognosis. HTLV-1 can infect a variety of cell types, but more than 90\% of infected cells are CD4+ memory T cells in vivo \cite{ATL_review}. In order to specifically separate HTLV-1-infected cells from other normal T-cells, Cell adhesion molecule 1 (CADM1/TSLC1) is used because of its sensitivity and specificity \cite{cadm1,cadm1_2}. Thus, in this study, to purify leukemic cells (HTLV-1 infected cells) from the peripheral blood mononuclear cells (PBMC) of ATL patients, we used the cell surface markers shown in Table \ref{Table2}.
}

\ak{The objective of our analysis using leukemic samples was to evaluate
which type of hematopoietic cell is closest to a given leukemic sample at the chromatin level.
Specifically,} we added the ATAC-seq data of a leukemic sample to healthy hematopoietic ATAC-seq data
and calculated the Hamming distances where \rr{$(\Mc^*,p_G^*)=(64000,10^{-2})$ is used}. 
We computed $\mathcal{H}_{\mathrm{Ward}}(\zeta,\mathbb{S}_\nu)$
as the distance between cell type $\nu\in\mathbb{T}$ and leukemic sample $\zeta$;
in this case, sample $\zeta$ was extracted from one patient.

We define the $q$-th closest cell type of sample $\zeta$ as
type $\nu_\zeta^{(q)}\in\mathbb{T}$ \bl{to provide} the $q$th minimum
of $\mathcal{H}_{\mathrm{Ward}}(\zeta,\mathbb{S}_\nu)$ in terms of $\nu$.
Using this quantity, we define the rank gap between a given reference type $T_0 \in \mathbb{T}$
and sample $\zeta$ as
\[G_{T_0,\zeta}=q-1,
\] such that $T_0=\nu_\zeta^{(q)}$.
In particular, we call $\nu_\zeta^{(1)}$ the closest type of sample $\zeta$.
Note that rank gap $G_{T_0,\zeta}=0$ holds when $T_0=\nu_\zeta^{(1)}$.
Thus, we not only revealed the closest cell type, but also identified the second, third, and so on closest cell type,
and quantified the difference between the characterization results of our algorithm and a given type as the rank gap.

\blue{As shown in Table \ref{Table3}, by calculating the Hamming distance between each CLL sample and a set of hematopoietic
  cells, we found that the closest cell type for all CLL samples was B cells,
 which coincides well with the characteristics of CLL cell surface markers. 
 This result led us to conjecture that our method could infer the cell type of a given leukemic cell
 characterized by immunophenotypes with using only its ATAC-seq data.}

\begin{table}[htb]
  \begin{tabular}{|c|c|c|c|} \hline
    &&Type & ``closest cell type" \\
    sample name ($\zeta$)& SRR number & consistent to  & calculated by \\
    &&  surface marker   & our algorithm ($\nu_\zeta^{(1)}$)\\ \hline
    CLL1 & SRR6762820& B & B \\
    CLL2 & SRR6762844& B & B \\
    CLL3 & SRR6762861& B & B \\
    CLL4 & SRR6762895& B & B \\
    CLL5 & SRR6762925& B & B \\
    CLL6 & SRR6762952& B & B \\
    CLL7 & SRR6762968& B & B \\
    \hline
  \end{tabular}
  \caption{
{\bf Classification of ATAC-seq data of CLL samples.}
    ``Closest cell type'' computed by our algorithm.}
   \label{Table3}
\end{table}

\blue{In order to assess the applicability of our method to leukemia whose cell of origin is not uniform and
  has high levels of heterogeneity between cases, we analyzed AML samples \cite{AML}.
 We found that the results of our analysis \rr{for pHSC and Blast} had substantial overlap
 with those of a previous study \cite{AML}\rr{, where 12 out of 16 samples for pHSC and 13 out of 18 samples for Blast
 are overlapped,} as shown in Table \ref{Table4}.
However, in the case of LSC, we found differences between the results of our analysis and those from \cite{AML}.
Most of the LSC samples were closest to LMPP using our algorithm, but to GMP in \cite{AML}.
As mentioned above, the LSC used in the present study were CD99-positive and are presumed to be composed of LMPP-like cells,
which suggests that our characterization by using information of the Hamming distance
infers the cell type with high accuracy, though further investigation is required.}
 \begin{table}[htb]
 \centering
  \begin{tabular}{|c|c|c|c|c|} \hline
    &&``closest normal cell" ($T_0$)& ``closest cell type" &\\
    sample name ($\zeta$)& SRR number& calculated in Fig 6i & calculated by&rank gap \\
    && from Ref. \cite{AML}  & our algorithm ($\nu_\zeta^{(1)}$) & ($G_{T_0,\zeta}$)\\ \hline
    SU654-pHSC & SRR2920595& MPP & MPP & 0\\
    SU353-pHSC & SRR2920571& MPP & MPP & 0\\
    SU351-pHSC &SRR2920568& MPP & LMPP & 1\\
    SU209-pHSC1& SRR2920564& GMP & MPP & 4\\
    SU209-pHSC2 &  SRR2920562& GMP & GMP & 0\\
    SU209-pHSC3 & SRR2920561& GMP & GMP & 0\\
    SU070-pHSC1 & SRR2920557& HSC & MPP & 1\\
    SU070-pHSC2 &SRR2920556&  HSC & HSC & 0\\
    SU048-pHSC & SRR2920552& MPP & MPP & 0\\
    SU583-pHSC1 & SRR2920588& GMP & LMPP & 2\\
    SU583-pHSC2 & SRR2920587& GMP & GMP & 0\\
    SU575-pHSC  & SRR2920584& MPP & MPP & 0\\
    SU501-pHSC  & SRR2920581& MPP & MPP & 0\\
    SU496-pHSC  & SRR2920579& MPP & MPP & 0\\
    SU484-pHSC & SRR2920576& MPP & MPP & 0\\
    SU444-pHSC & SRR2920574& MPP & MPP & 0\\
    \hline
    SU654-LSC & SRR2920594& LMPP & LMPP  & 0\\
    SU583-LSC & SRR2920586& GMP & LMPP & 1\\
    SU575-LSC & SRR2920583& GMP & LMPP & 2\\
    SU496-LSC  & SRR2920578& GMP & GMP & 0\\
    SU444-LSC  & SRR2920573& GMP & LMPP & 1\\
    SU353-LSC & SRR2920570& GMP & LMPP & 1\\
    SU209-LSC  &SRR2920559& GMP & LMPP & 1\\
    SU070-LSC  & SRR2920555& GMP & LMPP & 1\\
    \hline
    SU654-Blast & SRR2920593& GMP & LMPP & 1\\
    SU444-Blast & SRR2920572& Mono & Mono & 0\\
    SU353-Blast & SRR2920569& GMP & GMP & 0\\
    SU351-Blast & SRR2920567& Mono & GMP & 1\\
    SU209-Blast & SRR2920558& GMP & GMP & 0\\
    SU070-Blast1 & SRR2920554& Mono & Mono & 0\\
    SU070-Blast2 & SRR2920553& Mono & Mono & 0\\
    SU048-Blast1 & SRR2920551& GMP & GMP & 0\\
    SU048-Blast2 & SRR2920550& GMP & Mono & 1\\
    SU048-Blast3 & SRR2920549& GMP & GMP & 0\\
    SU048-Blast4 & SRR2920548& GMP & Mono & 1\\
    SU048-Blast5  & SRR2920547& GMP & GMP & 0\\
    SU048-Blast6 & SRR2920546& GMP & GMP & 0\\
    SU583-Blast & SRR2920585& GMP & GMP & 0\\
    SU575-Blast & SRR2920582& GMP & LMPP & 1\\
    SU501-Blast & SRR2920580& Mono & Mono & 0\\
    SU496-Blast & SRR2920577& GMP & GMP & 0\\
    SU484-Blast & SRR2920575& Mono & Mono & 0\\
    \hline
  \end{tabular}
  \caption{
{\bf Classification of ATAC-seq data of AML samples.}
    Comparison between the ``closest normal cell'' in Fig 6i of \cite{AML}
    and ``closest cell type'' computed by our algorithm.
    The second, the third, and $\cdots$-th ``closest type'' were also identified by our algorithm.
    The rank gap represents the difference of the result between the two analytical methods.
    For example, the ``closest normal cell'' of sample SU351-pHSC is MPP in \cite{AML},
    but is LMPP by our algorithm. MPP was the second ``closest cell type''.
    Thus, the rank gap was calculated as 2-1 (=1).
    If the results from the two analytical methods coincide with each other, the rank gap is 0.}
   \label{Table4}
\end{table}

\begin{figure}[htbp]
\includegraphics[width=12cm,clip]{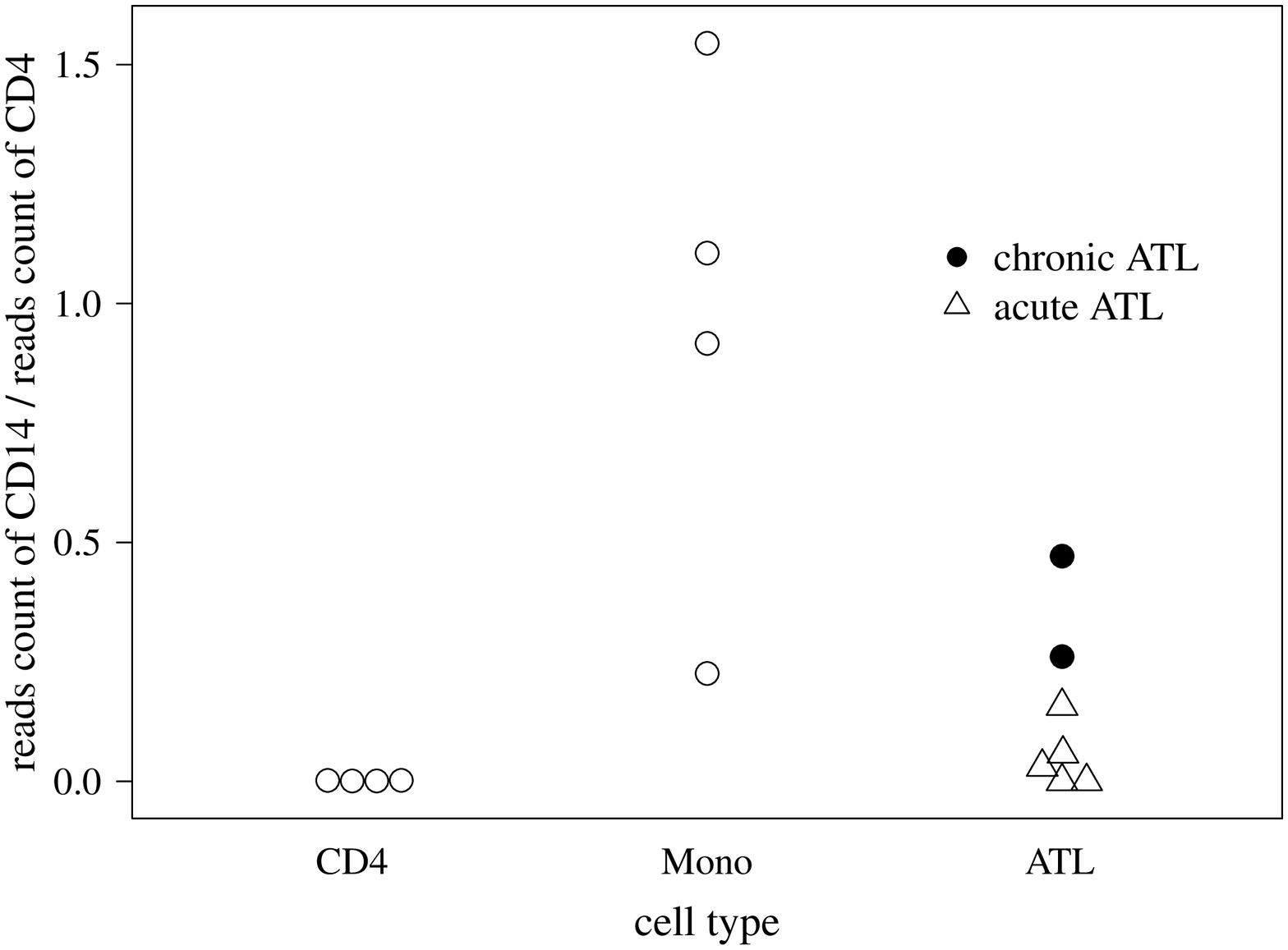}
  \caption{
{\bf Comparison of RNA-seq data among CD4$^+$T, Mono and ATL samples.}
    The reads count of CD14 over the reads count of CD4 from RNA-seq data of
    CD4$^+$T, Mono, and ATL samples.}
\label{Figure15}
\end{figure}

 Finally we analyzed ATL samples (See Materials and methods for details of sample preparation).
 When we calculated the Hamming distance between each ATL sample and a set of hematopoietic cells,
we found that the closest cell type for two ATL samples was Mono (hereafter we term these samples "Mono-like ATL"),
while that of the other samples was CD4$^+$T, as shown in Table \ref{Table5}.
Surprisingly, the two Mono-like ATL samples were categorized into chronic-type ATL.
\blue{Since CD14 is the marker of Mono (Table \ref{Table1}), we investigated the CD14 gene expression pattern in 
CD4$^+$T, Mono and ATL samples.
Particularly, we calculated the ratio of the CD14 reads count to the CD4 reads count from RNA-seq data
and found that the two Mono-like ATL samples exhibited higher values among all ATL samples (Fig \ref{Figure15}).}
\blue{In this way,
  the obtained results led us to conjecture that our algorithm could infer the cell phenotype,
  potentially including clinical subtypes, only using ATAC-seq data.
However, we need to analyze more samples to validate this conclusion.}

 \begin{table}[htb]
  \begin{tabular}{|c|c|c|c|} \hline
    & & & ``closest cell type'' \\
    sample name ($\zeta$)&  DRR number&clinical subtypes   &calculated by our algorithm ($\nu_\zeta^{(1)}$) \\ \hline
    ATL1 & DRR250710&Acute &   CD4$^+$T \\
    ATL2 &  DRR250711&Acute &  CD4$^+$T \\
    ATL3 &  DRR250712&Acute &  CD4$^+$T \\
    ATL4 &  DRR250713&Acute & CD4$^+$T \\
    ATL5 &  DRR250714&Acute &  CD4$^+$T \\
    ATL6 & DRR250715&Chronic &   Mono \\
    ATL7 & DRR250716& Chronic &  Mono \\
    \hline
  \end{tabular}
  \caption{
{\bf Classification of ATAC-seq data of ATL samples.}
    Clinical subtypes of ATL samples
    and ``closest cell type'' computed by our algorithm. }
   \label{Table5}
\end{table}





\section{Discussion}
In this paper, we presented a new algorithm to \rr{systematically perform clustering of} epigenomic data using the Hamming distance,
which enabled us to find optimal parameters of the data reduction \rr{toward cell-type classification}.
\ak{This algorithm has one clear advantage in terms of computational cost
  compared to \rrr{a previous method} using targeted regions merged over samples \rrr{\cite{AML}}.
  Especially, when adding new samples to the analysis, we only have to calculate the distances
  between newly appearing pairs of samples and not between preexisting samples.
  \rr{The computational cost of} the presented systematic algorithm is \rr{significantly lower} for this situation
  \rr{compared to the \rrr{previous method} with merging targeted regions.}
Furthermore, this algorithm was found to effectively detect the closest cell type of a leukemic sample,
with the results being broadly consistent with the characterization of leukemic samples by cell surface markers or RNA-seq.
Thus, the developed algorithm potentially serves as a screening for the phenotype of a leukemia sample
by using the ATAC-seq data of the sample as input.}

\rrr{As a next step, we need to investigate if our constructed algorithm is robust for other existing methods and data.
For example, for the same data of hematopoietic cells,
we replaced the Hamming distance with the Dice coefficient, which has been used in the CODEX project {\cite{codex}} 
to quantify the differences between two samples,
but found the results with $p_G=10^{-2}$ were not improved in terms of the penalty. 
We also compared our algorithm with DiffBind {\cite{diffbind}}, which is commonly used as a ChIP-seq differential analysis tool, 
but again found that DiffBind with its default setting did not give a better clustering result. Note that there are other existing methods and data to be checked in the future.
}

A unique point of our constructed algorithm is that we only used ATAC-seq data without gene expression data.
Our analysis suggests that ATAC-seq data itself contains enough information to determine cell types even
in the absence of regional annotation data such as promoters or enhancers.
\ak{This feature implies that} our algorithm reveals elusive epigenomic properties
that significantly affect the phenotype of cell types.
Another advantage of our algorithm is that
we do not assume a strong property for the statistics of the reads data, which is otherwise implicitly assumed when quantile normalization is performed.
Instead of using the strong assumption, we took a data-driven approach for the normalization of the reads data,
where we pre-analyzed the statistics of the reads data before performing any normalization.

Finally, our algorithm could extend its application to leukemic samples whose properties are uncertain.
We also expect that our \re{whole} approach with slight modifications will be applicable to other epigenetic sequencing data
such as ChIP-seq and bisulfite sequencing \rrr{available, for example, from The International Human Epigenome Consortium (https://epigenomesportal.ca/ihec/), ROADMAP Epigenomics (http://www.roadmapepigenomics.org/) and many other resources,} whose target regions for the analysis are not uniform between samples.

\section{Materials and methods}

\subsection{Ethics Statement}
Experiments using clinical samples were conducted according to the principles expressed in the Declaration of Helsinki and approved by the Institutional Review Board of Kyoto University (permit numbers G310 and G204). ATL patients provided written informed consent for the collection of samples and subsequent analysis. 

\subsection{Sequencing sample preparation}
\blue{ATL patient PBMCs were thawed and washed with PBS containing 0.1\% BSA. To discriminate dead cells, we used the LIVE/DEAD Fixable Dead Cell Stain Kit (Invitrogen). For cell surface staining, cells were stained with APC anti-human CD4 (clone: RPA-T4) (BioLegend) and anti-SynCAM (TSLC1/CADM1) mAb-FITC (MBL)
 antibodies for 30 minutes at 4\(^\circ\)C followed by a wash with PBS. HTLV-1 infected cells (CADM1+ and CD4+) were sort-purified with FACS Canto (Beckman Coulter) to reach 98-99$\%$ purity. Data was analyzed by FlowJo software (Treestar).
 Soon after the sorting, 10000-50000 HTLV-1 infected cells were centrifuged and used for ATAC-seq as previously described \cite{ATAC1}. Total RNA was isolated from the remaining cells using the RNeasy Mini Kit (Qiagen). Library preparation and high-throughput sequencing were performed by Macrogen Inc. (Seoul, Korea). The diagnostic criteria and classification of clinical subtypes of ATL were performed as previously described \cite{shimoyama}. 77 ATAC-seq datasets from 13 human primary blood cell types and datasets from 42 AML patients were obtained from the Gene Expression Omnibus (GEO) with accession number GSE74912 \cite{AML}. ATAC-seq datasets from 7 CLL patients were obtained from GSE111015 \cite{CLL2020} and RNA-seq datasets of CD4$^+$T and Mono cells
 were obtained from GSE74246 \cite{AML}.}

\subsection{Sequencing data analysis}
\blue{ATAC-seq reads were aligned using BWA version 0.7.16a \cite{BWAMEM} with default parameters. SAMtools \cite{sam} was used to convert SAM files to compressed BAM files and sort the BAM files by chromosome coordinates.
PICARD software (v1.119) (\url{http://broadinstitute.github.io/picard/})
was then used to remove PCR duplicates using the MarkDuplicates options.
Reads with mapping quality scores less than 30 were removed from the BAM files.
For peak calling, MACS2 (v2.1.2) software was used \cite{MACS2}.
RNA-seq data were aligned to human reference genome hg19 using STAR 2.6.0c \cite{star}
with the {-}{-}quantMode GeneCounts function. Normalization was not performed, and only raw reads count data of CD14 and CD4 were used in this study.}

\subsection{Principles of data reduction}\label{PAlgo}
When \rd{we analyze preprocessed ATAC-seq data with $\hat{\mathbf{P}}$,
  we have to care for biases caused by the fact that the amount of reads, $\NumR$,
  depends on the setting of the sample preparation and on the sequencers used.
  \re{(See Appendix for the explicit construction of $\hat{\mathbf{P}}$.)}
Normalization is done to remove such biases.}

A conventional way to perform normalization is to use quantile normalization,
where the distribution of the reads number on certain regions in the DNA is assumed to be the same for all samples \cite{quantile, quantile2}.
However, there is no strong reason to support \rd{this assumption, particularly for sample sets of different cell types.}
    \rd{Furthermore, under} this assumption, there is a risk that we overlook
important differences between different cell types. \rd{Therefore, in this paper}, we do not assume this property.

An alternative way to perform normalization
is to reduce \rd{the} data into a simple binary value $h_{\gamma,x}$ \gr{$\in\{0,1\}$}
on each genomic position $(\gamma,x)$, where $h_{\gamma,x}$ depends on the data size $\NumR$ as \rd{little} as possible.
For example, one could determine the state of $h_{\gamma,x}=1$ and $h_{\gamma,x}=0$
as \rd{an} ``open'' and ``closed'' chromatin status, respectively, on genomic position $(\gamma,x)$. 

In this direction, our ultimate purpose is to look for the ``best'' principle
\rd{that determines} two states for $h_{\gamma,x}$, by which a set of samples including different cell types
are completely classified into groups of the same cell type.
We \rd{use no information about cell types when determining the value of $h_{\gamma,x}$,
because we would like to have an algorithm that can be applied without knowing the cell types.}

\subsection{Peak-calling with ranking}\label{Callpeak}
Currently we do not have the best solution to properly
determine \rd{two effective states }for $h_{\gamma,x}$.
\gr{As a candidate to approach the best solution},
 we use the MACS2 algorithm, which was originally invented to analyze ChIP-seq data \cite{MACS2}
but is now widely used to estimate the location of open chromatin regions from ATAC-seq data \cite{Greenleaf,Howard}.

We would like to find the set of position $(\gamma, x)$
where the number of reads overlapping with position $(\gamma, x)$,
$Y_{\gamma, x}(\hat{\mathbf{P}})$, is relatively high in \rd{the neighborhood $(\gamma, x)$.}
The MACS2 algorithm is likely to detect \rd{those positions} from the data of the reads described by $\hat{\mathbf{P}}$. 
In our calculation, we use the MACS2 (v2.1.2) callpeak command with option
\textsf{``{-}{-}nomodel {-}{-}nolambda {-}{-}keep-dup all -p $p_G$''},
where we need to set parameter $p_G$ as a parameter of {\it peak} inference
(for details, see \cite{MACS2}).

\rd{By applying MACS2 to the input ATAC-seq data, we obtain the following output data structure}:
\begin{itemize}
	\item The label $\gamma_k \in \Chr$ of the chromosome to which the $k$-th peak has
	  a start position $1 \le \alpha_k \le L_\gamma$ and end position
          $1 \le \beta_k \le L_\gamma$ for $1 \le k \le M$
	(here $M$ is the number of peaks). 
	We call $g_k = (\gamma_k, \alpha_k, \beta_k)$ the \textit{$k$-th peak region}.
      \item For each $g_k$, $p$-value $p_k$ with \rd{$p_k\le p_G$ is associated to the $k$-th peak}.
        Note that MACS2 \rd{outputs $\log_{10}(1/p_k)=-{\log_{10}p_k}$ instead of $p_k$.}
\end{itemize}
$\Chr$ and $L_\gamma$ are the set of all chromosomes
and the length of chromosome $\gamma$, respectively \re{(see Appendix for details of the notations)}.
\rd{We define $\mathbf{A}$ as}
\begin{align*}
  \mathbf{A} := (g_k,p_k)_{k=1}^M,\\
g_k := (\gamma_k, \alpha_k, \beta_k).
\end{align*}
By reordering \rd{the terms of $k$}, we can set $p_k\le p_{k'}$ for any $k < k'$ without loss of information. 

In Fig \ref{Figure2}, we show the distribution of the peak width $|\beta_k-\alpha_k|$ versus ranking $k$.
Note that $g_k$ with \rd{high $p_k$} could be affected significantly by the conditions of the experiments including sequencing,
because the data above \bl{rank value} $40000$ unnaturally touches the value of the lower limit of width $200$,
\rd{which is predetermined by the MACS2 algorithm.}
Thus, there is a possibility that peaks with higher $p$-values
could strongly depend on both the inference algorithm and the number of reads $\NumR$.
Those peaks would presumably not contribute to \rd{the detection of }cell phenotypes.
This observation suggests \rd{we should} remove peaks with higher $p$-values as mentioned in Results.

\subsection{\gr{Parameterized} binarization by cutting off low-ranked peaks} \label{Binary}
\rd{Next we reconsidered} how to alleviate biases in the data
by introducing threshold number $\Mc$, such that
\[\overline{\mathbf{A}}(\Mc):=\{g_k\}_{k=1}^{\Mc},
\]
\rd{which leads to the removal of $\{g_k\}_{k=\Mc+1}^M$ as a candidate for the normalization of the ATAC-seq data.}
Note that $\overline{\mathbf{A}}(\Mc=\infty)=\{g_k\ |\ (g_k,p_k)\in{\mathbf{A}} \}$.
Then, by using $\overline{\mathbf{A}}$, we may introduce a binary sequence
\[
\mathbf{B}:=\{\hx\}_{\gamma\in\Chr,1\le x \le L_\gamma},
\] such that $\hx = 1$ if there is $k$ satisfying $\alpha_k\le x\le \beta_k$
with $(\alpha_k,\beta_k)\in \overline{\mathbf{A}}$; otherwise $\hx=0$ as shown in Fig \ref{Figure4}.

$p_G$ and $\Mc$ can be regarded as parameters \rd{for 
determining the value of $\hx$ within the MACS2 algorithm and
what part of the data} is taken into account, respectively.
Thus, our task under the principle above turns out to be how to determine a proper set of ($\Mc$,$p_G$)
\rr{for the cell-type classification.}

\subsection{Hamming distance} \label{Hamming}
\textit{The Hamming distance} is often used to compare two binary sequences
in information theory (see Section 13 in \cite{Hamming})
  and is equal to the number of positions on which two symbols have different values.
  \rdyi{See Fig \ref{Figure3} for an illustrative explanation.}

The Hamming distance between two binary sequences \rd{$\BSeq^{c_1}$ and $\BSeq^{c_2}$} with $c_1,c_2\in\mathbb{S}$ is defined as
\[
	H(\mathbf{B}^{c_1}, \mathbf{B}^{c_2}) := \sum_{\substack{\gamma \in \Chr \\ 1 \le x \le L_\gamma}} 
	\delta(\hx^{c_1}, \hx^{c_2}),
\]
where we define
\[
	\delta(\hx^{c_1}, \hx^{c_2}) = \begin{cases}
		1 & (\hx^{c_1} \neq \hx^{c_2}) \\
		0 & (\hx^{c_1} = \hx^{c_2}).
	\end{cases}
\]

\subsection{\rdyi{Algorithm of hierarchical clustering}}\label{UPGMA}

\rdyi{In this and the next subsection, 
we recall algorithms for agglomerative hierarchical clusterings and drawing dendrograms.
We use two methods, \textit{UPGMA} and \textit{Ward's}.
Though they are described in many textbooks (for example, see Chapter 4 in \cite{UPGMA}), 
we need the description in order to define the global penalty and the type penalty.
Our description of the algorithms follows \cite{Hier}. 
}

To describe the algorithms, we define \rdyi{two distance functions} 
between two subsets, $\freeSet_1, \freeSet_2 \subset \Sample$ \rdyi{as follows
  (for inductive definitions and other distance functions, see Section 4.2 in \cite{UPGMA}).
One distance function, $\mathcal{H}_{\mathrm{UPGMA}}$
comes from the UPGMA method and is defined as}
the average of all the distances between \rd{samples in $\freeSet_1$ and $\freeSet_2$.}
Equivalently, we define
\[
	\rdyi{\mathcal{H}_{\mathrm{UPGMA}}}(\freeSet_1, \freeSet_2) := \frac{1}{|\freeSet_1||\freeSet_2|} 
	\sum_{c_1 \in \freeSet_1} \sum_{c_2 \in \freeSet_2} H(\BSeq^{c_1}, \BSeq^{c_2}).
\]
If $\freeSet_1$ or $\freeSet_2$ is empty, we set $\rdyi{\mathcal{H}_{\mathrm{UPGMA}}}(\freeSet_1, \freeSet_2) = 0$.

\rdyi{Another choice of the distance function, $\mathcal{H}_{\mathrm{Ward}}$, 
comes from Ward's method and is defined as
\[
	\mathcal{H}_{\mathrm{Ward}}(\freeSet_1, \freeSet_2) := 
	\sqrt{\frac{D_{1,2}}{|\freeSet_1| + |\freeSet_2|}-\frac{|\freeSet_2|D_1}{|\freeSet_1|(|\freeSet_1| + |\freeSet_2|)}
	 - \frac{|\freeSet_1|D_2}{|\freeSet_2|(|\freeSet_1| + |\freeSet_2|)}}
\]
where we define
\begin{align*}
	D_{1} &:= \frac{1}{2}\sum_{c_1 \in \freeSet_1}\sum_{c_2 \in \freeSet_1} H(\BSeq^{c_1}, \BSeq^{c_2})^2, \\
	D_{2} &:= \frac{1}{2}\sum_{c_1 \in \freeSet_2} \sum_{c_2 \in \freeSet_2} H(\BSeq^{c_1}, \BSeq^{c_2})^2, \\
	D_{1,2} &:= \sum_{c_1 \in \freeSet_1}\sum_{c_2 \in \freeSet_2} H(\BSeq^{c_1}, \BSeq^{c_2})^2.
\end{align*}
Again, if $\freeSet_1$ or $\freeSet_2$ is empty, we set $\rdyi{\mathcal{H}_{\mathrm{Ward}}}(\freeSet_1, \freeSet_2) = 0$.
}

In the following, we fix $\mathcal{H}(\freeSet_1, \freeSet_2)$ as $\mathcal{H}_{\mathrm{UPGMA}}$ or $\mathcal{H}_{\mathrm{Ward}}$.
We sometimes identify sample $c \in \Sample$ and subset $\{c\}$ of single element $c$. 
For example, we write $\mathcal{H}(\freeSet_1, c_2)$ for $\mathcal{H}(\freeSet_1, \{c_2\})$. 
Note that $\mathcal{H}(\{c_1\}, \{c_2\}) = \mathcal{H}(c_1, c_2) = K H(\BSeq^{c_1}, \BSeq^{c_2})$
where $K=1$ for $\mathcal{H}=\mathcal{H}_{\mathrm{UPGMA}}$ and $K=2^{-1/2}$ for $\mathcal{H}=\mathcal{H}_{\mathrm{Ward}}$ by definition.

We define a \textit{cluster} as subset $\Cluster$ of $\Sample$
with a specified order of elements.  
\rd{Hierarchical clustering is an algorithm that can construct set $\Memory_{\NumS}$ of clusters
and order the elements in $\Sample$ to draw dendrograms.}

\begin{enumerate}
	\item We set $\Cluster_{\tau} := \{\tau\}$ for $1 \le \tau \le \NumS$. 
	We do not consider the order of the elements in $\Cluster_\tau$
	because they are sets of a single element.
	\item We define the list of uncombined clusters as $\List_{1} := \{\Cluster_1, \Cluster_2, \dots, \Cluster_{\NumS}\}$
	and set the historical list of clusters as $\Memory_{1} = \List_{1}$.
	\item 
	At the $t$-th step $(1 \le t \le \NumS - 1)$, we define $\Cluster_{t + \NumS}, \List_{t + 1}$ and $\Memory_{t+1}$ inductively.
	\begin{enumerate}
		\item We look up the pair $\Cluster_{\tau'}$ and $\Cluster_{\tau''}$ with $\tau'<\tau''$ in $\List_{t}$
    	such that their distance is \gr{a minimum}; that is,
    	\[
    		\mathcal{H}(\Cluster_{\tau'}, \Cluster_{\tau''}) = 
    		\min_{\substack{\Cluster',\Cluster'' \in \List_{t}\\\Cluster'\ne\Cluster''}} \mathcal{H}(\Cluster', \Cluster'').
    	\]
    	Note that $1\le \tau'<\tau'' < t+\NumS$ by construction.
		We consider only the case when the pair is uniquely determined.
    	\item 
    	We define a new cluster $\Cluster_{t + \NumS} = \Cluster_{\tau'} \cup \Cluster_{\tau''}$.
    	If the elements of $\Cluster_{\tau'}$ are ordered as $c_1, c_2, \dots, c_z$ and 
    	the elements of $\Cluster_{\tau''}$ are $c'_1, c'_2, \dots, c'_{z'}$, then
		the elements of $\Cluster_{t + \NumS}$ \rd{are} ordered as
    	\[
    		c_1, c_2, \dots, c_z, c'_1, c'_2, \dots, c'_{z'}.
    	\]
    	\item We define
		\begin{align*}
			\List_{t+1} &:= (\List_{t} \setminus \{\Cluster_{\tau'}, \Cluster_{\tau''}\}) \cup \{\Cluster_{t+\NumS}\}, \\
			\Memory_{t+1} &:= \Memory_{t} \cup \{\Cluster_{t + \NumS}\}.
		\end{align*}
		If $t < \NumS - 1$, go to the $(t+1)$-th step.
	\end{enumerate}
\end{enumerate}

We can easily see that if we do not consider the ordering, then we have $\Cluster_{2\NumS-1} = \Sample$ as a set.
Thus we finally obtain a list of $2\NumS-1$ clusters $\Memory_{\NumS} = \{ \Cluster_1, \Cluster_2, \dots, \Cluster_{2\NumS-1}\}$ 
and an ordering of all elements of $\Sample$ from $\Cluster_{2\NumS-1}$.

\subsection{How to draw dendrograms}
The \textit{(rooted) dendrogram} displays how our clustering combines pairs of clusters and the distance of the pairs. In the following, we explain an algorithm that introduces new symbols. \rdyi{For details, see \cite{Hier}.}

\begin{enumerate}
	\item If sample $\tau \in \Sample$ appears in the ordering of $\Cluster_{2\NumS-1}$ as the $a_\tau$-th element,
	then we associate point $n_\tau = (a_\tau, 0)$ in two-dimensional coordinate space to cluster $\Cluster_{\tau}$.
	\rd{We call point $n_\tau$ the \textit{leaf}, which corresponds to $\Cluster_\tau$.}
	\item For $1 \le t \le \NumS - 1$, we inductively associate point $n_{t+\NumS}$ to cluster $\Cluster_{t+\NumS}$.
	  If $\Cluster_{t+\NumS}$ is constructed as the union of $\Cluster_{\tau'}$ and $\Cluster_{\tau''}$
          with $1\le \tau'<\tau'' < t+\NumS$, 
	we associate to $\Cluster_{t + \NumS}$ the node
	\[
		n_{t+\NumS} = \left( a_{t + \NumS} = \frac{a_{\tau'} + a_{\tau''}}{2}, \mathcal{H}(\Cluster_{\tau'}, \Cluster_{\tau''}) \right).
	\]
	Note that $\Cluster_{\tau'}$ and $\Cluster_{\tau''}$ are uniquely determined.
	We call $n_{t+\NumS}$ the \textit{node} associated to the $(t+\NumS)$-th cluster $\Cluster_{t+\NumS}$.
	\item We connect $n_{t+\NumS}$ with $n_{\tau'}$ and $n_{\tau''}$. 
\end{enumerate}

Since each node or leaf $n$ corresponds to cluster $\Cluster$, we can define the \textit{offspring set} $\mathcal{B}_n$ of $n$ as set $\Cluster$ without ordering.
Graphically, the offspring set of node $n$ is the set of samples corresponding to leaves branched from node $n$, as displayed in Fig \ref{Figure7}.
This intuitional explanation is justified, since the $y$-coordinate of the ``mother node'' $n_{t+\NumS}$ is larger than or equal to those of the ``child nodes'' $n_{\tau'}, n_{\tau''}$ if we use Ward's method or UPGMA.
Note that there are many choices to draw dendrograms; for example, 
at any branching node, we can exchange two branches without any essential change in the data structure.

\subsection{Global penalty as a cost function}\label{Penalty}
In this section, we discuss the global penalty,
a quantity that measures how the obtained hierarchical clustering differs
from our knowledge of \rd{cell type classifications.}
We also give examples displaying the computation of the penalties and extreme situations
\rd{that represent the theoretical bounds of the penalties.}
Note that these examples are just for explanation and \textit{not} obtained from actual data.

In our settings, each sample is previously classified by \textit{types}.
Explicitly, set $\Types$ consists of thirteen types:
\[
\Types = \{\Bcell, \CDfour, \CDeight, \CLP, \CMP, \Ery, \GMP, 
	\HSC, \LMPP, \MEP, \Mono, \MPP, \NK\}.
\]
For each type $\nu \in \Types$, we denote the set of samples classified to type $\nu$ as $\Sample_\nu$.
\rd{This set} could be empty, though it is not in our case.
For every pair $\nu, \nu'$ of distinct types,
there are no common elements in $\Sample_\nu$ and $\Sample_{\nu'}$, and the union of $\Sample_\nu$
among all types $\nu \in \Types$ coincides with $\Sample$. Equivalently, 
\[
	\Sample = \bigcup_{\nu \in \Types} \Sample_\nu.
\]

For a given hierarchical clustering constructed in the manner of the previous section,
the \textit{type penalty} for type $\nu$ is the quantity $\lambda_{\nu}$ defined as follows.
If $\Sample_{\nu}$ is empty, we set $\lambda_\nu = 0$. Otherwise, since the cluster grows step by step,
there is the minimum $\gr{\tau}$ for $1 \le \tau \le 2\NumS - 1$ such that  $\Sample_{\nu} \subset \Cluster_{\tau}$.
We denote the minimum $\tau$ by $\tau(\nu)$.  
Then we define $\lambda_\nu$ as the number of elements in $\Cluster_{\tau(\nu)}$ \rd{that are not} of type $\nu$.
In other words, we set
\[
	\lambda_{\nu} := |\Cluster_{\tau(\nu)}| - |\Sample_{\nu}|.
\]
Since $\Cluster_\tau$ includes all elements of type $\nu$, we find $\lambda_\nu \ge 0$. 
Also since $\Cluster_\tau$ is a subset of $\Sample$, we find $\lambda_{\nu} \le |\Sample| - |\Sample_{\nu}|$.
Thus we have
\[
	0 \le \lambda_{\nu} \le |\Sample| - |\Sample_{\nu}|.
\]
(See Fig\ \ref{Figure7} for an explanation of type penalties.)

\begin{figure}[t]
\includegraphics[width=18cm,clip]{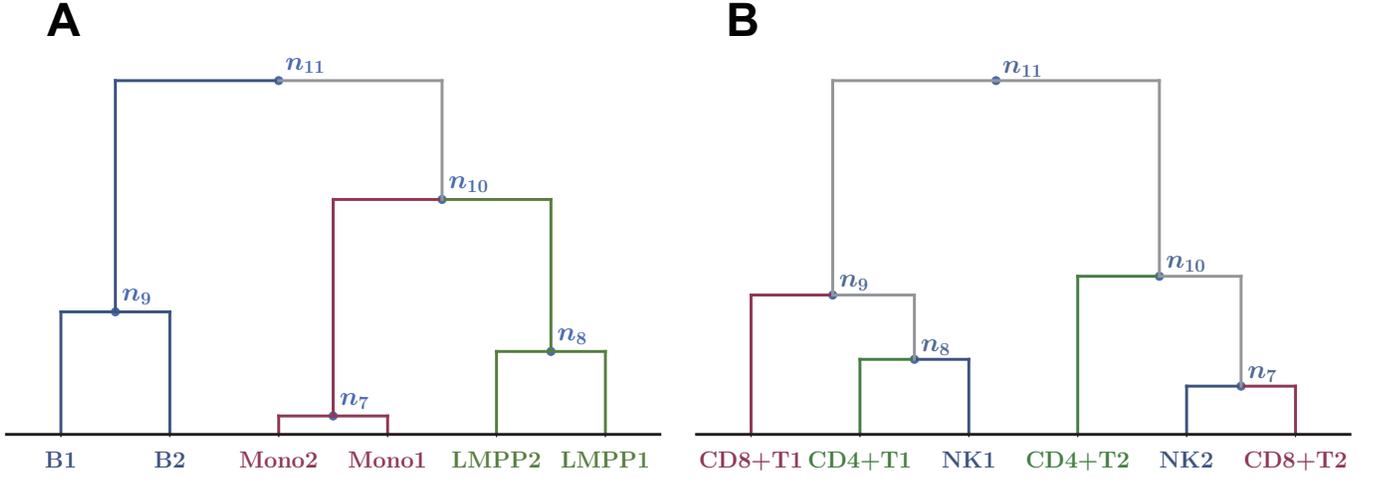}
 \caption{
{\bf Examples of dendrograms with extreme penalties.}
Note that this dendrogram is constructed using artificial data to explain how to calculate the penalty, though we use the same labels such as Mono1.
Both of these dendrograms have six leaves ($|\Sample| = 6$)
that are classified into three types (in these examples, $|\Types| = 3$).
(A) \rd{This} example gives the lowest global penalty 0.
(B) \rd{In this} example, we have $\tau(\CDfour) = \tau(\CDeight) = \tau(\NK) = 11$.
Since the corresponding cluster $\Cluster_{11}$ is the whole set $\Sample$,
the local penalty is $6-2=4$ for each type, and the global penalty is $4 \times 3 = 12$.
This result \bl{gives} the upper bound $(|\Types| - 1) \cdot |\Sample| = (3-1) \cdot 6 = 12$ in \rd{equation} \eqref{Eq:Estim}.}
\label{Figure16}
\end{figure}

For a given hierarchical clustering, the \textit{global penalty} $\lambda$ is defined to be the total sum of type penalties,
\[
	\lambda := \sum_{\nu \in \Types} \lambda_\nu.
\]
\bl{$\lambda$} is bounded as
\begin{equation}\label{Eq:Estim}
	0 \le \lambda \le \sum_{\nu \in \Types} (|\Sample| - |\Sample_{\nu}|) = (|\Types| - 1) \cdot |\Sample|.
\end{equation}
In our case, since $|\Types| = 13$ and $|\Sample| = 77$,
we have $0 \le \lambda \le (13-1) \cdot 77 = 924$.
Note that for a certain class of trees, these upper and lower bounds are not achieved.
Fig \ref{Figure16} displays examples of the upper and lower bounds.

Further, we write $\lambda(\Mc,p_G)$ as $\lambda$ to point out that $\lambda$ depends on $(\Mc,p_G)$.
\re{Note that $\Cluster_{\tau(\nu)}$ is equal to $\mathcal{B}_{n_{\tau(\nu)}}$, which was defined in the previous section.}

\section*{Acknowledgments}
The authors thank P. Karagiannis for valuable comments and proofreading of this manuscript.
They also thank MACS Program at Graduate School of Science Kyoto University which allowed this collaboration to be carried out.


%
%

\newpage
\appendix

\section{ATAC-seq: Analysis for open chromatin regions based on Tn5-transposase} \label{ATAQ}
Throughout this appendix, we used hg19 as the human reference sequence. 
\rd{It consists} of $24$ groups of symbol sequences, which corresponds to chromosomes labeled
\rd{as} $\Chr := \{1,2,\dots, 22, \chrX, \chrY\}$.
\rd{The} underlying structure of a chromosome is a long chain of DNA and
the DNA is represented as \gr{a sequence of elements in set}
\[
\DNA := \{\mathrm{A, T, G, C}\},
\]
where each symbol corresponds to \rd{the nucleotides}
adenine (A), thymine (T), guanine (G), and cytosine (C).

For the $\gamma$-th chromosome ($\gamma \in \Chr$),
the length of the corresponding DNA sequence is written as $L_{\gamma}$,
where $5.0 \times 10^7 \le L_\gamma \le  2.5 \times 10^8$ 
and the total length is $L = \sum_{\gamma=1}^{22}L_\gamma + L_{\chrX} + L_{\chrY} \sim 3.1 \times 10^9$.
\rd{To position the} $x$-th base pair in the $\gamma$-th chromosome,
we \bl{set}chromatin
\[
b^{\gamma}_{x} \in \DNA,
\]
with $1\le x\le L_\gamma$. 
In this paper, for a set $\mathbb{SET}$, we write the number of elements in $\mathbb{SET}$ as $|\mathbb{SET}|$.
For example, we have $|\DNA| = 4$ and $|\Chr| = 24$.

Chromatin is a complex of DNA and associated proteins such as histones.
A chromatin has ``open'' regions, around which the density of \rd{the} DNA and the associated proteins
are rather low and also ``closed'' regions, around which the opposite situation happens.
Gene \rd{expressions are} largely regulated by the \rd{interactions between DNA} and transcription factors depending on \bl{the open and closed regions}.
  The analysis of open/closed chromatin regions is necessary for the understanding
  of cell differentiation and phenotype \cite{Epigenome,chromatin1}.

  ATAC-seq was developed for the genome-wide detection of open chromatin regions.
  One of the features of ATAC-seq is that it uses Tn5 transposase.
At a certain proper condition, Tn5-transposase mainly cut DNA in open chromatin regions
and the sequences of those DNA fragments are obtained by sequencers \cite{ATAC1}.
ATAC-seq has several advantages compared to the other epigenomic sequencing methods \cite{sequence}.
For example, \rd{to analyze open chromatin regions, DNase-seq needs about $10^7$--$10^8$ cells and takes $4$--$5$ days to obtain the data.}
ATAC-seq, on the other hand, requires only about $10^3$--$10^4$ cells and takes half a day.

\section{Reads}\label{Reads}
As briefly reviewed above, one Tn5-transposase cuts \rd{and splits DNA into two parts or fragments.}
If there are two Tn5-transposases, two locations of DNA are cut \rd{to make three fragments.}

Thus, \rd{we can view} \textit{fragment} $f$ as a subsequence of
a DNA sequence consisting of successive symbols in $\mathbb{D}$.
Since we refer to the same DNA sequence of the human genome in this study,
fragment $f$ can be also represented by three coordinates:
the \rd{the chromosome number} $\gamma \in \Chr$,
the start position $s = s(f)$, and the end position $e = e(f)$, where $1\le s \le e\le L_\gamma$.
In other words, $f$ is \bl{a sequence} $(b^{\gamma}_{s}, b^{\gamma}_{s+1}, \dots, b^{\gamma}_{e})$, 
\bl{that} can be \rd{expressed as $f = (\gamma, s, e)$.}

A \textit{sample} is, in our settings,
a product generated by a certain experimental procedure through ATAC-seq library preparation from a set of cells \cite{ATAC1}.

The input of a sequencer is the set of the obtained fragments
$\left\{f_i\right\}_{i=1}^{\NumF}$, where \rd{a fragment $f_i$ is $( \gamma_i, s_i, e_i)$,}
its length $L(f_i)$ is equal to $e_i-s_i+1$, and the number of fragments is denoted as $\NumF$. 
A sequencer with ``paired-end sequencing'' outputs a DNA sequence of the two edges of a fragment as two reads
$(\Reads_i^s,\Reads_i^e)$ where
\begin{eqnarray*}
  \Reads_i^s:=(\ra_j)_{j=1}^{\ell_i},{\ }\Reads_i^e:=(\ra_j')_{j=1}^{\ell_i} {\rm\ for\  } \ra_j,\ra_j'\in\mathbb{D},
\end{eqnarray*}
 meaning that \bl{each} length of \rd{the} two reads $(\Reads^s_i,\Reads^e_i)$ is $\ell_i$.

We consider \textit{read} \gr{as} a sequence of four symbols in $\DNA$ of length less than or equal to $\ell_{0}$,
where $\ell_{0}$ can be changed as a parameter controlled by the sequencer.
Note that for the case of ``single-end sequencing'', where one gets only a read from one edge,
we obtain read $\Reads_i=\{\ra_j\}_{j=1}^{\ell_i}$.
\rd{In the end}, we obtain the data of reads $\Reads := \{\Reads_i\}_{i=1}^{\NumR'}$ where the number of reads is denoted as $\NumR'$.
Note that in the case of ``paired-end sequencing'', one may regard both $\Reads_i^s$ and $\Reads_i^e$ as $\Reads_i$.
This is the starting point of our analysis because sequencers do not directly give the actual values of $f_i$.

Summarizing the relationship between fragments and reads,
let us assume that all reads are obtained from ``paired-end sequencing''
and \rd{that} the sample preparation and the sequencer output \rd{are} ``ideal'' as follows. 
If we denote fragment $f_i$ as sequence $(b^{\gamma_i}_{s_i}, b^{\gamma_i}_{s_i+1}, \dots, b^{\gamma_i}_{e_i})$,
then the beginning read $\Reads_i^s$ and the terminal read $\Reads_i^e$ corresponding to $f_i$ are
\begin{align*}
	\Reads_i^s &= \begin{cases}
		(b^{\gamma_i}_{s_i}, b^{\gamma_i}_{s_i + 1}, \dots, b^{\gamma_i}_{s_i + \ell_{0}-1}) & {\rm for\ } \ell_{0} \le L(f_i), \\
		(b^{\gamma_i}_{s_i}, b^{\gamma_i}_{s_i + 1}, \dots, b^{\gamma_i}_{e_i}) & {\rm for\ } \ell_{0} > L(f_i),
	\end{cases} \\
	\Reads_i^e &= \begin{cases}
		(b^{\gamma_i}_{e_i-\ell_{0}+1}, b^{\gamma_i}_{e_i - \ell_{0}+2}, \dots, b^{\gamma_i}_{e_i}) & {\rm for\ } \ell_{0} \le L(f_i), \\
		(b^{\gamma_i}_{s_i}, b^{\gamma_i}_{s_i + 1}, \dots, b^{\gamma_i}_{e_i}) & {\rm for\ } \ell_{0} > L(f_i).
	\end{cases}
\end{align*}

In other words, if the length $L(f_i)$ of fragment $f_i$ is greater than or equal to $\ell_{0}$, 
the beginning read $\Reads_i^s$ is the direct inference of the first $\ell_{0}$ symbols of the fragment $f_i$.
\rd{The condition for terminal read $\Reads_i^e$ is similar.}
If length $L(f_i)$ is less than $\ell_{0}$,
we directly infer $\Reads_i^s = \Reads_i^e = f_i$ as a sequence of four symbols,
where we see that the two reads have the same length.

However, the situation above is ``ideal'' and \rd{there are unexpected errors that }stochastically flip symbols in the ideal situation.
Thus, we need to infer the information of fragments in a statistical manner.
Note that this inference can be straightforwardly applied to the case of ``single-end sequencing''.

\begin{figure}
\begin{center}
\includegraphics[width=13cm,clip]{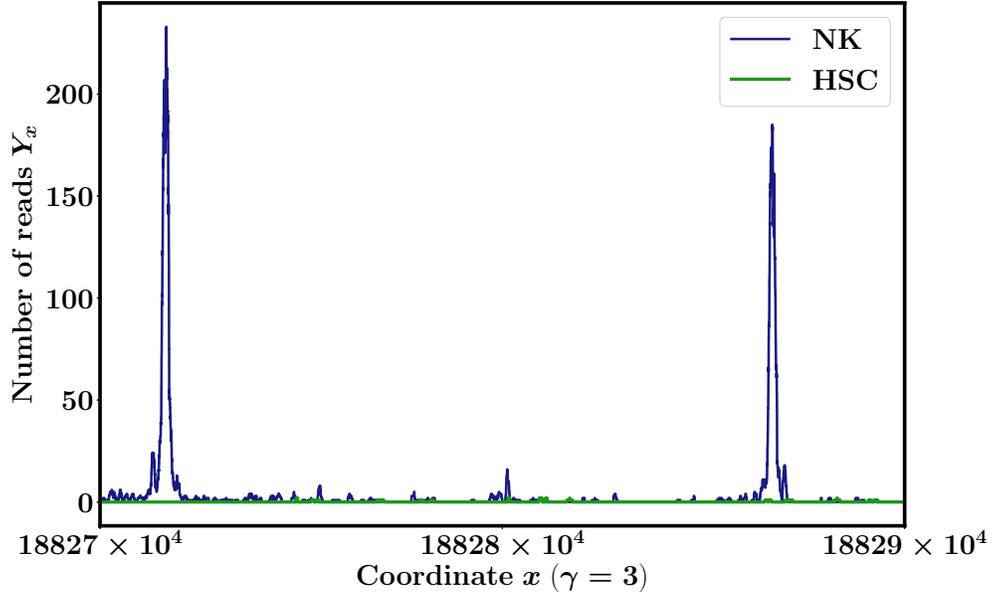}
\caption{{\bf The number of reads vs genomic position.}
  The number $Y_x = Y_{\gamma,x}$  of reads in \rd{the ATAC-seq data (vertical axis)
 vs position $x$ in the DNA sequence (horizontal axis),}
  where $x$ starts \rd{from $18827\times10^4$ and ends at $18829\times10^4$}, and chromosome number $\gamma=3$.}
\label{S1Fig}
\end{center}
\end{figure}

\section{Alignment of reads onto the reference genome}\label{Alignment}
Hereafter, for simplicity, we consider single-ended reads $\Reads=\{\Reads_i\}_{i=1}^{\NumR'}$
because similar processes can be done for \bl{paired-end} reads.
We perform mapping of the reads data $\Reads$ from a sequencer onto the DNA sequence. 

We use \rd{the} \textsf{BWA-MEM} algorithm of the software \textit{BWA} (v0.7.16a) with no options.
\rd{This algorithm} aligns each read onto the hg19 reference sequence $(b^{\gamma}_{x})_{\gamma \in \Chr, 1 \le  x \le L_\gamma}$ 
and gives an estimate of the quality of the alignment (for details, see \cite{BWAMEM} and references therein).
Then we obtain the following data:
\begin{itemize}
\item \rd{Chromosome number $\estim{\gamma}(\Reads_i)\in \Chr\cup\{\rm U\}$ with the start position $\estim{s}(\Reads_i)$
  and the end position $\estim{e}(\Reads_i)$ of read $\Reads_i$ mapped onto the DNA sequence,}
  where $1\le \estim{s}(\Reads_i)\le\estim{e}(\Reads_i)\le L_{\estim{\gamma}(\Reads_i)}$.
  Note that \rd{${\rm U}$ is a set of unplaced sequences in any elements in $\Chr$.}
  \rd{Hereafter, $\Chr$ includes ${\rm U}$} with \gr{$L_{\mathrm U}\simeq 3.7\times10^6$}.
	\item The mapping quality score $\MQ(\Reads_i)\ge 0$ of read $\Reads_i$ \bl{calculated} by \rd{using the} Phred quality score.	
\end{itemize}

\rd{Therefore,} $(\estim{\gamma}(\Reads_i),\estim{s}(\Reads_i), \estim{e}(\Reads_i))$
infers the coordinates
$(\gamma_i, s_i, e_i)=(\estim{\gamma}(\Reads_i), \estim{s}(\Reads_i), \estim{e}(\Reads_i))$
of read $\Reads_i$ \rd{onto} the DNA sequence.
\bl{For $\Reads_i$, we define $\hat{\mathbf{T}}(\Reads_i)$ as}

\[
\hat{\mathbf{T}}(\Reads_i):=(\estim{\gamma}(\Reads_i), \estim{s}(\Reads_i),\estim{e}(\Reads_i), \MQ(\Reads_i)).
\]

To select reliable data with $\hat{\mathbf{T}}(\Reads_i)$, we preprocess the outputs obtained above as follows:
\begin{enumerate}
\item In order to reduce duplicated reads, which could be produced artificially in the sequence sample preparation,
  we apply the command \textsf{MarkDuplicates} \rd{in} PICARD software (v1.119) (http://broadinstitute.github.io/picard/)
	with the \textsf{REMOVE\_DUPLICATE} option.
	\item Then we cut off reads with a mapping quality score $\MQ(\Reads_i)$ less than 30. 
	We used \textit{samtools} \rd{for this purpose} \cite{sam}. 
\end{enumerate} 

After processes (1) and (2), we obtain
\[
\hat{\mathbf{P}}(\Reads'):= \{ (\estim{\gamma}(\Reads_i'), \estim{s}(\Reads_i'), \estim{e}(\Reads_i') ) \}_{i=1}^{\NumR},
\]
where $\ell_i'$ is the length of $\Reads_i'$ and $\NumR$ denotes the number of reads after preprocessing.
$\{\Reads_i'\}_{i=1}^{\NumR}$ can be straightforwardly determined by $\{\Reads_i\}_{i=1}^{\NumR'}$.
This is part of \rd{the} information obtained by the preprocessing.
Note that $\MQ(\Reads_i') \ge 30$ holds for any $i$ with $1\le i \le \NumR$
and there are no duplicated pairs in $\hat{\mathbf{P}}$.
For simplicity, hereafter, we sometimes express $\hat{\mathbf{P}}(\Reads)$ as $\hat{\mathbf{P}}$.
\rd{We use similar abbreviations for other symbols.}

\section{Pilings of reads}\label{Piling}
From the data $\hat{\mathbf{P}}$, we can calculate how many reads are on position $(\gamma, x)$ in the DNA sequence.
We consider the set of reads located on position $(\gamma, x)$
symbolically \rd{by defining}
\[
\mathbb{Y}_{\gamma, x}(\hat{\mathbf{P}}(\Reads)) :=
\left\{
1 \le i \le \NumR \; \mid \; 
\estim{\gamma}(\Reads_i) = \gamma \ {\rm and} \
( \; \estim{s}(\Reads_i) \le x \le \estim{e}(\Reads_i) )
\right\}.
\]
For two samples in reads data $\Reads$ obtained from SRA (SRR2920495.sra and SRR2920466.sra),
we show $Y_{\gamma, x} := |\mathbb{Y}_{\gamma, x}|$,
which is the number of reads on each position $(\gamma,x)$ in the DNA sequence in 
Fig \ref{S1Fig}.
In this study, we used reads data $\Reads$ from the Gene Expression Omnibus (GEO)
with accession number GSE74912 as the initial input of the analysis.

\end{document}